\documentclass[preprint]{aastex63}
\usepackage{natbib}
\usepackage{url}
\usepackage{latexsym}
\usepackage{etoolbox}
\usepackage{float}

\newcommand{\teff}{\ensuremath{T_{\rm eff}}}

\newcommand{\acen}{$\alpha$ Cen}
\newcommand{\mearth}{M$_\oplus$}
\newcommand{\rearth}{R$_\oplus$}

\shorttitle{\acen\ A Planets}
\shortauthors{Beichman et al.}
\received{Sept. 1, 2019}

\begin{document}
\title{Searching for Planets Orbiting \acen\ A with the James Webb Space Telescope}
\author[0000-0002-5627-5471]{Charles Beichman}
\affiliation{NASA Exoplanet Science Institute\\
Infrared Processing and Analysis Center\\
Jet Propulsion Laboratory\\ California Institute of Technology, Pasadena CA 91125}
\author[0000-0001-7591-2731]{Marie Ygouf}
\affiliation{NASA Exoplanet Science Institute, IPAC, Pasadena, CA 91125}
\author{Jorge Llop Sayson}
\affiliation{California Institute of Technology, Pasadena, CA 91125}
\author[0000-0002-8895-4735]{Dimitri Mawet}
\affiliation{California Institute of Technology, Pasadena, CA 91125}
\author[0000-0002-4263-2562]{Yuk Yung}
\affiliation{California Institute of Technology, Pasadena, CA 91125}
\author[0000-0002-9173-0740]{Elodie Choquet}
\affiliation{Aix Marseille Univ, CNRS, CNES, LAM, Marseille, France }
\author[0000-0003-0626-1749]{Pierre Kervella}
\affiliation{LESIA, Observatoire de Paris, Universit\'e PSL, CNRS, Sorbonne Universit\'e, Univ.~Paris Diderot, Sorbonne Paris Cit\'e, Paris, France}
\author[0000-0001-9353-2724]{Anthony Boccaletti}
\affiliation{LESIA, Observatoire de Paris, Université PSL, CNRS, Sorbonne Université, Université de Paris, 5 place Jules Janssen, 92195 Meudon, France}
\author{Ruslan Belikov}
\affiliation{NASA Ames Research Center, Space Science and Astrobiology Division, MS 245-3, Moffett Field, CA 94035, USA}
\author[0000-0001-6513-1659]{Jack J. Lissauer}
\affiliation{NASA Ames Research Center, Space Science and Astrobiology Division, MS 245-3, Moffett Field, CA 94035, USA}
\author{Billy Quarles}
\affiliation{Center for Relativistic Astrophysics, School of Physics, Georgia Institute of Technology, 
Atlanta, GA 30332, USA}
\author{Pierre-Olivier Lagage}
\affiliation{AIM, CEA, CNRS, Université Paris-Saclay, Université Paris Diderot,\\ Sorbonne Paris Cité, F-91191 Gif-sur-Yvette, France}
\author{Daniel Dicken}
\affiliation{AIM, CEA, CNRS, Université Paris-Saclay, Université Paris Diderot,\\ Sorbonne Paris Cité, F-91191 Gif-sur-Yvette, France}
\author[0000-0003-2215-8485]{Renyu  Hu}
\affiliation{Jet Propulsion Laboratory, California Institute of Technology, Pasadena CA 91125}
\author[0000-0003-4205-4800]{Bertrand Mennesson}
\affiliation{Jet Propulsion Laboratory, California Institute of Technology, Pasadena CA 91125}
\author{Mike Ressler}
\affiliation{Jet Propulsion Laboratory, California Institute of Technology, Pasadena CA 91125}
\author{Eugene Serabyn}
\affiliation{Jet Propulsion Laboratory, California Institute of Technology, Pasadena CA 91125}
\author{John Krist}
\affiliation{Jet Propulsion Laboratory, California Institute of Technology, Pasadena CA 91125}
\author{Eduardo Bendek}
\affiliation{Jet Propulsion Laboratory, California Institute of Technology, Pasadena CA 91125}
\author[/0000-0002-0834-6140]{Jarron Leisenring}
\affiliation{Steward Observatory, University of Arizona, Tucson, AZ}
\author{Laurent Pueyo}
\affiliation{Space Telescope Science  Institute, Baltimore, MD}

\correspondingauthor{Charles Beichman}
\email{chas@ipac.caltech.edu}

\section{Abstract}

 $\alpha$ Centauri A  is the closest solar-type star to the Sun  and offers an excellent opportunity to detect the thermal emission of a mature planet heated by its host star. The MIRI coronagraph on the James Webb Space Telescope  (JWST) can search the 1-3 AU (1\arcsec-2\arcsec) region around \acen\ A  which is predicted to be stable within the \acen\ AB  system. We demonstrate that with reasonable performance of the telescope and instrument,   a  20 hr program combining    on-target and reference star observations at 15.5 $\mu$m could  detect thermal emission  from planets as small as  $\sim$5 R$_\oplus$. Multiple visits every 3-6 months would increase the geometrical completeness, provide astrometric confirmation of detected sources, and push the radius limit down to $\sim3$ R$_\oplus$. An exozodiacal cloud only a few times brighter than our own should also be detectable,  although a sufficiently bright cloud might obscure any planet present in the system. While current precision radial velocity (PRV)  observations set a limit of 50-100 \mearth\  at 1-3 AU for planets orbiting \acen\ A,  there is a broad range of exoplanet radii up to 10 R$_\oplus$ consistent with these mass limits. A carefully  planned observing sequence along with state-of-the-art post-processing analysis could reject the light from \acen\ A at the level of $\sim10^{-5}$ at 1\arcsec-2\arcsec\ and minimize the influence of \acen\ B located 7-8\arcsec\ away in the 2022-2023 timeframe.  These space-based observations would complement  on-going imaging experiments  at shorter wavelengths as well as PRV and astrometric experiments to detect planets dynamically. Planetary demographics  suggest that the likelihood of  directly imaging a  planet whose mass and orbit are  consistent with present PRV limits is small, $\sim$5\%, and possibly lower if the presence of a binary companion further reduces occurrence rates. However,  at a distance of just 1.34 pc, \acen\ A   is our closest sibling  star and certainly  merits close scrutiny.


\section{Introduction}

The detection, characterization, and   search for biomarkers in the atmospheres of Earth analogs in the Habitable Zones (HZ) of their host stars  are exciting goals of both ground- and space-based astronomy as described and prioritized in  the National Academy's Decadal Reviews \citep{Astro2010}, NASA's Strategic Plan\footnote{https://www.nasa.gov/sites/default/files/atoms/files/nasa\_2018\_strategic\_plan.pdf}, the Exoplanet Science Strategy Report\footnote{https://www.nap.edu/catalog/25187/exoplanet-science-strategy}~\citep{ExoplanetScienceStrategy2018}, and  the recently announced Breakthrough Initiative\footnote{https://breakthroughinitiatives.org/arewealone}. The  high degree of stellar rejection (10$^{-10}$ in the visible and 10$^{-7}$ in the thermal infrared) demanded to detect an Earth analog at  small angular separations, typically 10s of milliarcsec for most nearby solar type stars, represents a daunting challenge in both reflected visible light and emitted thermal radiation. Studies of observatories capable of achieving these levels have led to designs of  instruments for 30-40 m telescopes on the ground \citep{Kenworthy2016, Mawet2016, Skemer2018, Mazin2019}, 4 to 15  m   telescopes in space (Habex, \citet{Habex}; LUVOIR, \citet{LUVOIR}),  as well as earlier initiatives such as the  TPF-C coronagraph and   TPF-I/Darwin  mid-IR interferometer (\citet{Leger1996,Angel1997, Beichman2007}).  However,  by virtue of its proximity to the Sun, \acen\ A   offers an opportunity to use more modest and  more near-term facilities to  image  directly a mature planet ranging in size from Jovian-sized to Earth-sized. Proposals exist to use ground-based  8  m telescopes \citep{Kasper2017} or  a small visible telescope in space \citep{Belikov2015}.

\begin{figure} [h!]
\centering
\includegraphics[width=1\textwidth]{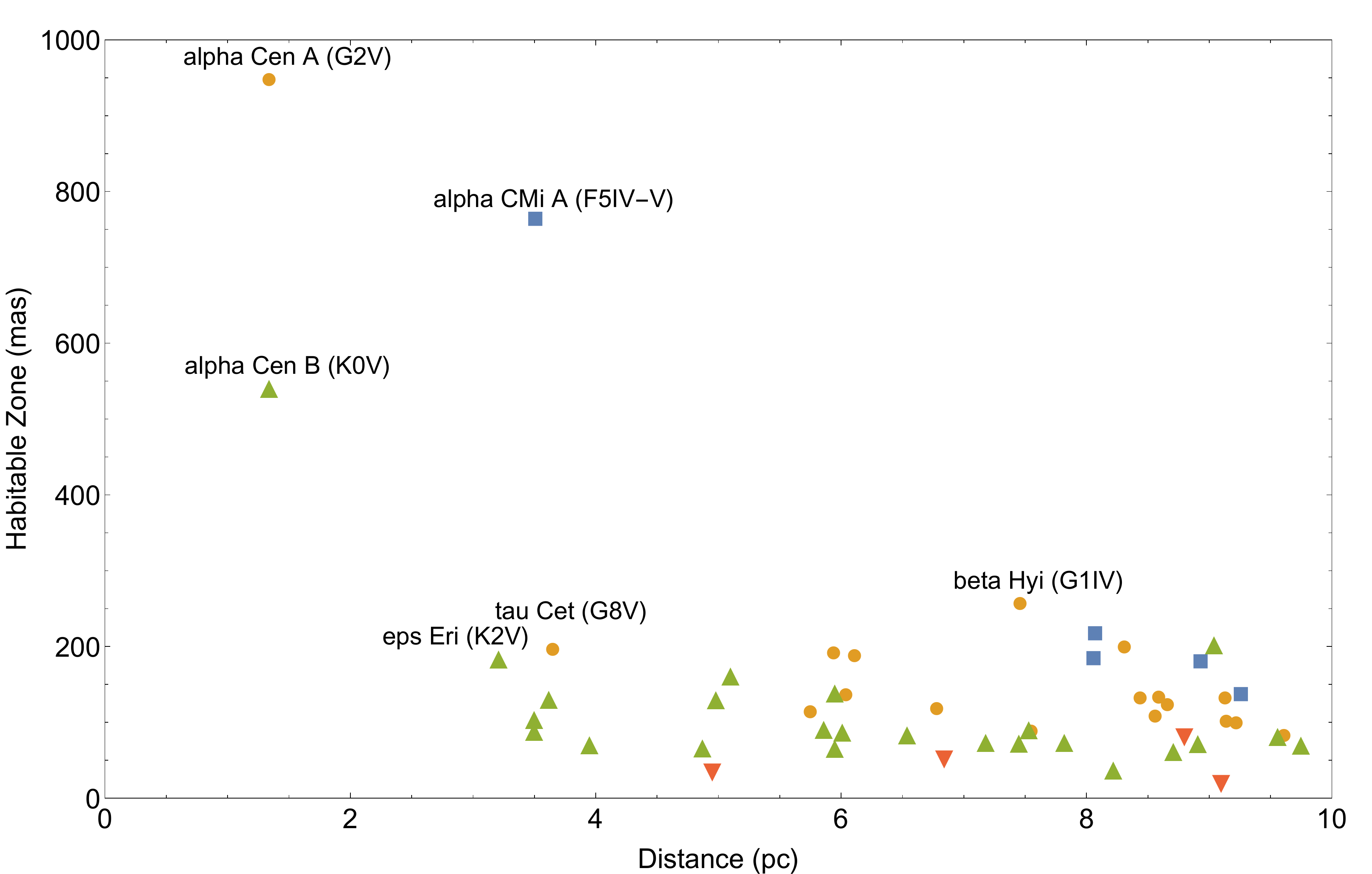}
\caption{ \acen\ A stands out as the most favorable star to examine due to the large  angular extent of its Habitable Zone, as indicated here as the angular separation (milliarcseconds, or mas) of a  planet receiving an Earth equivalent insolation from its host star \citep{Turnbull2015}. A few of the closest  and most prominent host stars are called out individually (F stars as blue squares, G stars as orange circles, K stars as green triangles, and M stars as inverted red triangles). \label{NearbyHZ}}
\end{figure} 

At a distance of 1.34 pc, \acen\ A is 2.7 times closer than the next most favorable G star, $\tau$ Ceti.  \acen\ A's luminosity of 1.5 L$_\odot$ \citep{Thevenin2002, Kervella2017} puts the center of its HZ (defined here as the separation of an Earth-Equivalent level of insolation, see also \citet{Kopparapu2017})   at a physical separation of 1.2 AU  which corresponds to an angular separation of 0.9\arcsec.   \acen\ A is the one stand-out exception, {\it primum ex parte},  in the list of solar-type host stars suitable  for the eventual detection and initial characterization of a HZ Earth (Figure~\ref{NearbyHZ}; \citet{Turnbull2015}). 

The 10-15 $\mu$m emission from \textit{an isolated   object}  with the same brightness as a warm  Earth-sized planet (20-40 $\mu$Jy)  would be   readily detectable by MIRI. There are, of course,  major challenges to be overcome:  the glare of \acen\ A, the presence of \acen\ B which might  remove planets from the \acen\ A system and which introduces a second source of noise, the stability of JWST and the performance of its coronagraphs. Yet these challenges can  be surmounted --- certainly for planets  larger than the Earth. We note that a search for planets orbiting \acen\ B is less favorable due to the  tight RV constraint on the presence of planets around \acen\ B \citep{Zhao2017}, its lower luminosity and correspondingly  smaller HZ ($\sim0.5$\arcsec), and to the  greater deleterious effects of  \acen\ A. 

\begin{figure}[h!]
\centering
\includegraphics[height=0.65\textwidth]{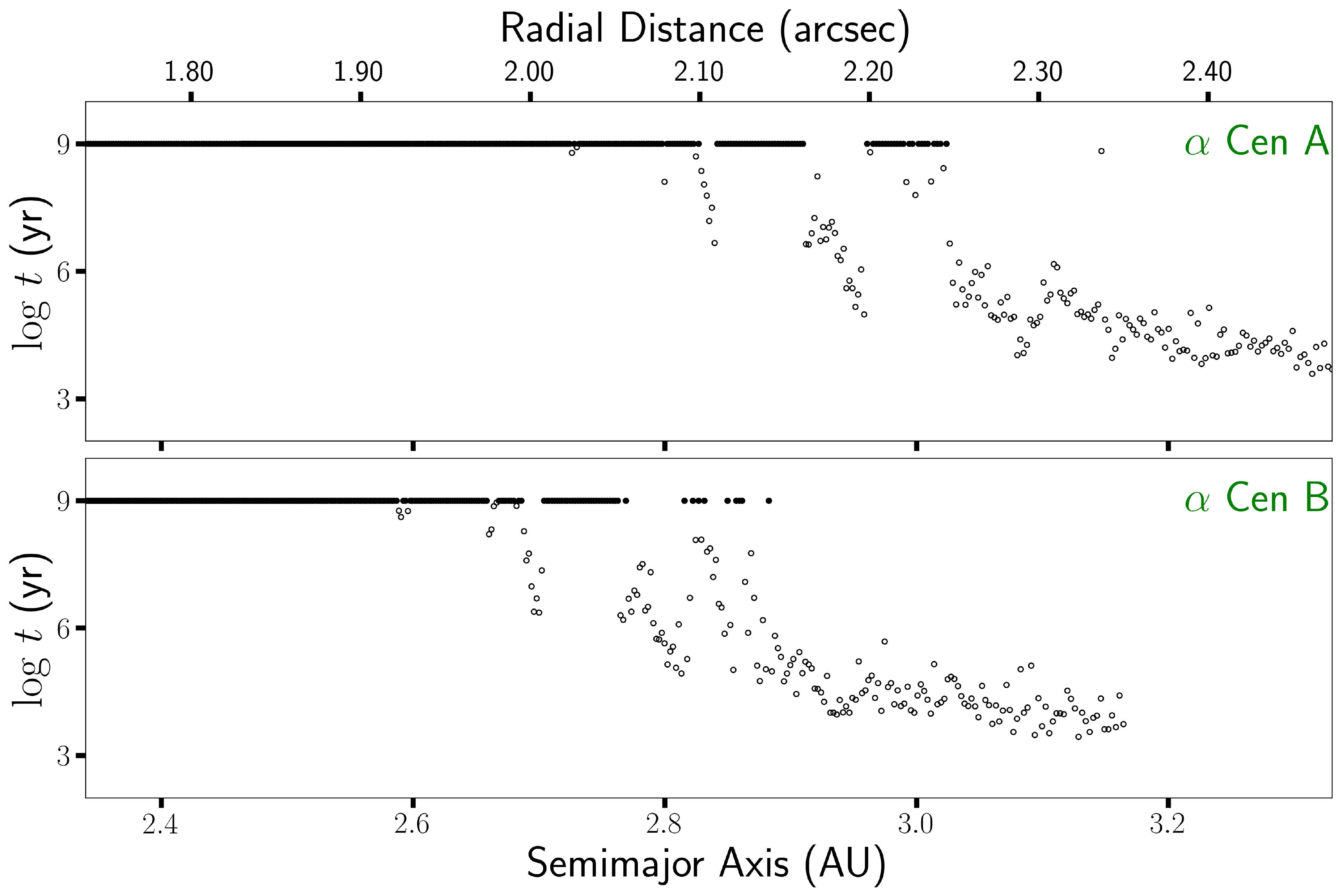}
\caption{Stable regions are found within $\lesssim$3 AU for planetary systems orbiting  \acen\ A and within $\sim$2.65 AU of \acen\ B \citep[based on work from][]{Quarles2018a}.\label{stability}}
\end{figure}

Current precision radial velocity (PRV) observations  \citep{Zhao2017}  constrain the mass of any planet near \acen\ A  to be $M\, sin(i)<$ 53  M$_\oplus$ in the Habitable Zone (1.2 AU).  Examination of their Figure 6 which includes their estimates for the effects of non-Gaussian noise sources (``red noise") suggests a limit between 50 and 100 \mearth ($2\, \sigma$). This limit applies to the near edge-on,  79$^o$, orientation of the \acen\ A-B system where dynamical studies indicate the presence of a stable zone $\lesssim$ 3 AU (or 2.1\arcsec) around \acen\ A despite the presence of \acen\ B (Figure~\ref{stability}; \citet{Quarles2016,Holman1999,Quarles2018a,Quarles2018b}).  There is a wide range of planet types  possible within these mass limits, from   Earth-sized planets to sub-Neptunes. In what follows we adopt an  upper limit to any radial velocity signature  of 5 m s$^{-1}$ which corresponds roughly to a 2$\sigma$ limit. We  recognize that future PRV observations will doubtless improve on this constraint.  Finally, we note that a planet's thermal emission depends  on its radius, not its mass, and the range of permissible {\it radii} is   broad due to wide range of observed planet densities. 

In this paper we investigate how a modest  observing program with the MIRI coronagraph  could  detect   HZ planets larger than $\sim$5  R$_\oplus$ orbiting \acen\ A  as well as a zodiacal dust cloud  only a few times brighter than our own cloud. Depending on the performance of JWST and the MIRI coronagraph, a more ambitious program could push to even lower planet sizes, $\sim$3  R$_\oplus$.

\section{The prospects for planets in the $\alpha$ Cen System \label{prospects}}
  
Statistical studies based on radial velocity (RV) and transit surveys help to assess whether  \acen\ A might  host one or more  planets. Unfortunately, transit surveys are incomplete in the 1-3 AU range for all radii \citep{Thompson2018} while RV data are incomplete for masses below 100 \mearth\ (Saturn) at these separations \citep{Cumming2008,Santerne2016}.   Combining various estimates suggests a cumulative planet incidence for FGK stars  of  3-8\% for M$>$ 100 M$_\oplus$ and P$<$5 yr with a five to tenfold increase for masses down to  10 M$_\oplus$. Thus, based on these statistical considerations there is a good chance (25--50\%) that \acen\ A might host one or more planets in the 10-100 \mearth\ range.

\citet{Fernandes2019} parameterize the joint planet occurrence rate as a function of period and mass:

\begin{equation}
     \frac{d^2N}{dlogPdlogM}=C_0f(P)(\frac{M}{10M_\oplus})^\gamma \,\,\,  
     \label{occurrence}
\end{equation}

\begin{equation} 
f(p)=\left(\frac{P}{P_{break}}\right)^{p1} \, \textrm{for} \,  P<P_{break}\,\,\,\textrm{; or}\,  \left(\frac{P}{P_{break}}\right)^{p2} \, \textrm{for} \, P \geq  P_{break}
\end{equation}

While there is considerable uncertainty in the fitted parameters, \citet{Fernandes2019} find that the following values provide a reasonable fit to the available data: $P_{break} = 1581 d,\, p1=-p2=0.65, \, \gamma=-0.45$ and $C_0$=0.84. If we integrate Eqn~\ref{occurrence} over  periods from  10 to 1800 days (corresponding to an outer limit of 3 AU) with a minimum mass of 10 \rearth\ and an upper mass consistent with an RV limit of 5 m s$^{-1}$, then \acen\ A has a $\sim$15\% probability of hosting  a planet with those properties. While the extrapolation to the lowest masses ($\sim$ 10 \mearth) is quite uncertain, the population estimates in the mass/radius range which we will show are accessible to JWST ($3\sim5$\rearth\ and P$<$1800 d,  $\S$\ref{monte}) are reasonably well-grounded in transit and RV data \citep{Cumming2008}.

One reason for pessimism about \acen\ A's suitability as a stellar parent comes from the fact that \acen\ A \& B form  a relatively tight binary system.  \citet{Kraus2016} have analyzed the statistics from Kepler transits and shown that detected planets are only about one-third as abundant in comparable-mass binary systems with projected separations of $<$ 50 AU as they are around single stars. However, \cite{Quintana2002} have shown that the late stages of planet growth for a prograde disk of planetary embryos and planetesimals orbiting about either \acen\ A or B near the plane of the binary orbit would grow into a configuration of terrestrial planets comparable to that formed by an analogous disk orbiting the Sun perturbed by Jupiter and Saturn.  \cite{Xie2010} and \cite{Zhang2018} have found favorable conditions for planetesimals to survive and grow to planetary embryos in disks with inclinations of up to 10$^\circ$ relative to the  binary orbit.  Simulations of \cite{Quarles2016} and references therein have shown that a planet can remain in a low-inclination, low-eccentricity prograde orbit for longer than the age of the system throughout the habitable zones of both \acen\ A and B. 

The  population studies mentioned above are given in terms of planet masses, whereas   JWST will detect thermal emission which depends on planet radius. For masses between 10-100\mearth, radii can range   from 2 to  10 \rearth\   \citep{Howard2013} with  dramatic effects on the photometric signal.  We will address the sample consistent with known occurrence rate, the RV limits and detectability by JWST in a subsequent section ($\S$\ref{monte}).

\begin{figure}[h!]
\begin{tabular}{c}
\includegraphics[width=0.8\textwidth]{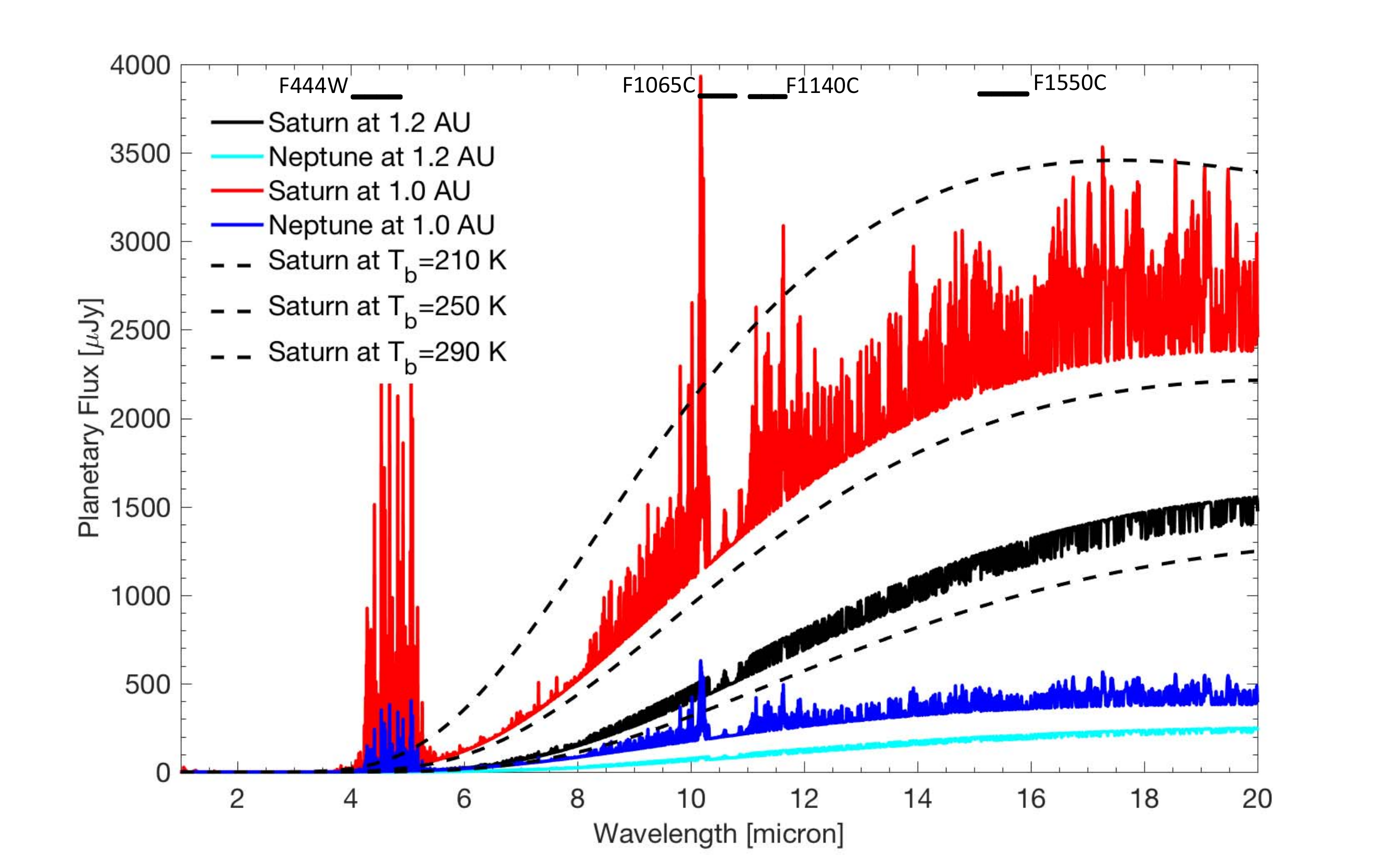}\\
\includegraphics[width=0.8\textwidth]{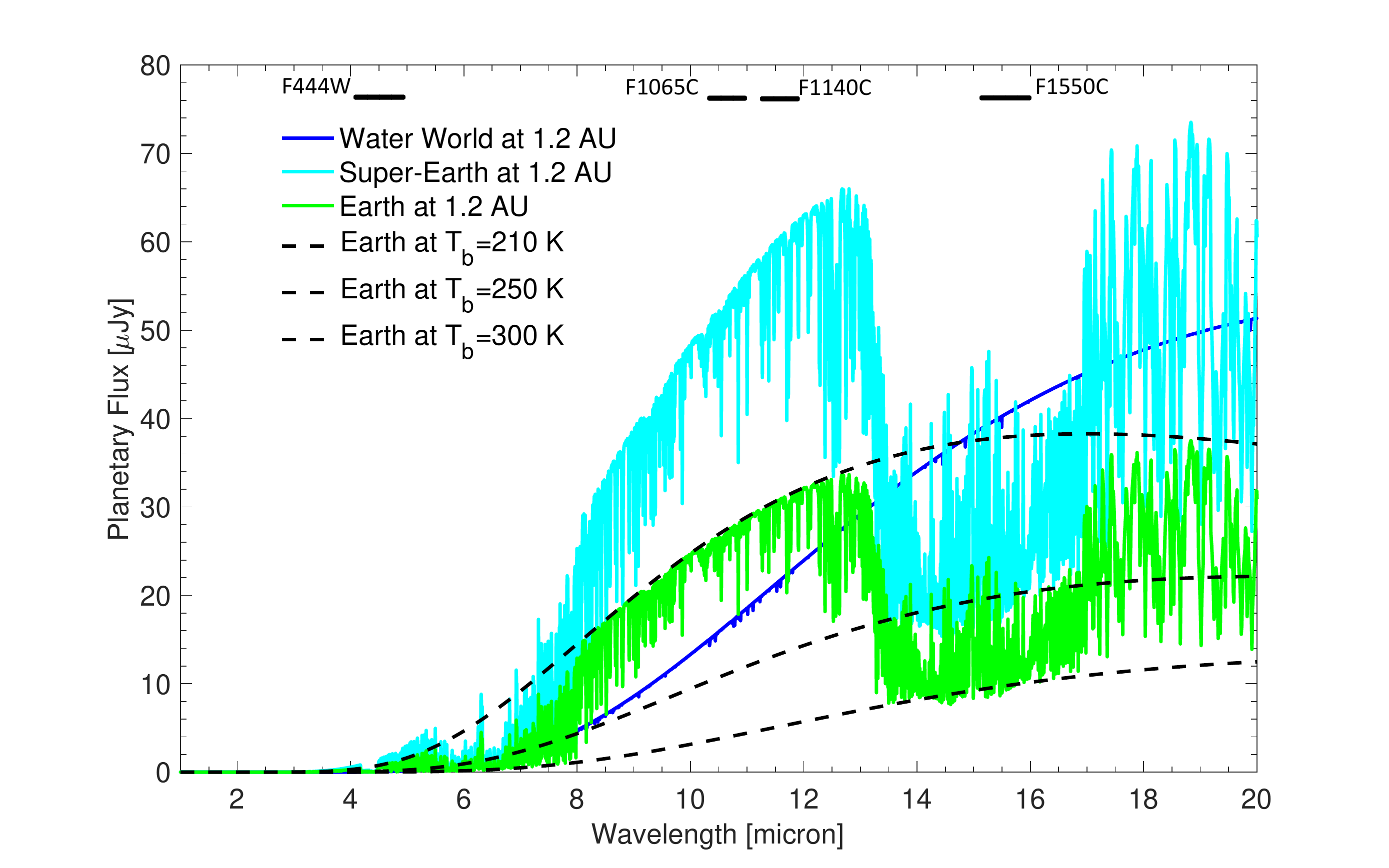}\\
\end{tabular}
\caption{The brightness of  a variety of model planets with radii between  4 to 10  $R_\oplus$  (a, top) and 1-2  $R_\oplus$ (b,bottom)  over a range of orbital locations and temperatures as described in Table~\ref{fluxes}. The locations of the 3 MIRI coronagraphic filters and one NIRCam filter are indicated.  \label{fig:flux}}
\end{figure}%

\section{Brightness of Habitable Zone Planets}

There is a broad base of literature  available to establish the expected level of brightness of exoplanets of various sizes and locations \citep{Burrows2004,Seager2010,Burrows2014}. We have developed a self-consistent  series of models based on the atmospheric chemistry and radiative transfer formalism  developed in Hu \& Seager (2013) and Hu (2014). Table~\ref{fluxes} and Figure~\ref{fig:flux} summarize models for  2, 4 and 10  R$_\oplus$  planets (mini-Neptunes, Neptunes, and Saturns) at 1.2 AU  with  a H$_2$-dominated atmosphere with   10$\times$  solar metallicity. The models include   condensation of water, ammonia, and methane when they reach saturation in the atmosphere and are thus suitable to simulate such  low-temperature atmospheres. As  water condenses in the atmosphere  water clouds form at a pressure of $\sim$0.1 bar. Due to the water clouds, the mid-infrared emission spectrum  is close to a 220 K black body as determined by the  cloud-top temperature. 

For planets  at 1.0 AU which receive about 50\% more irradiation than Earth our model predicts that water does not condense in its atmosphere and would likely be free of condensation clouds. The spectrum is dominated by strong H$_2$O, CH$_4$, and NH$_3$ absorption, as well as H$_2$-H$_2$ and H$_2$-He collision-induced absorption. There are  infrared windows into the hot, convective part of the atmosphere, at 4-5 $\mu$m, and to a lesser extent at 10 $\mu$m.

\begin{deluxetable}{ccccccc}
\tablecaption{ Predicted Brightness of Possible Planets Orbiting \acen\ A \label{fluxes}}
\tablehead{Planet	& Radius& Orbit&	T$_{eff}$ &	F1065C &F1550C &F$_{pl}$/F$_*$\\
Type& (R$_\oplus$) & (AU)& 	(K)& ($\mu$Jy)&	($\mu$Jy)&F1550C$^1$}
\startdata
Saturn      & 10& 1.2 &221  &500 &  1210&1.9$\times10^{-5}$ \\
Warm Saturn &10 & 1.0 &275  &1370& 2380&3.7$\times10^{-5}$\\
Neptune     &4  & 1.2 &221  &80 & 190&0.3$\times10^{-5}$\\
Warm Neptune &4 & 1.0 &262  &220&  380&0.6$\times10^{-5}$\\
Mini-Neptune& 2 & 1.2 &221  &20 & 50&0.08$\times10^{-5}$\\
Warm Mini-Neptune &2 &1.0&262&55&95&0.15$\times10^{-5}$\\ 
Water World &2  & 1.2 &215  &16 &  40&0.06$\times10^{-5}$\\
Super-Earth &1.4& 1.2 &250  &80&40&0.06$\times10^{-5}$\\
Earth       &1.0& 1.2 &250  &40 &  20&0.03$\times10^{-5}$\\ \hline
ExoZodi Cloud$^2$& -- &0.75 to 1.77  &250-300  &2500 &  3000&3.6$\times10^{-5}$\\ \hline
\enddata
\tablecomments{$^1$Contrast F(planet)/F(star). $^2$Estimated brightness of an analog of the Solar System's zodiacal cloud as discussed in $\S$\ref{HZdust}.}
\end{deluxetable}

\begin{figure}[h!] 
\centering
\includegraphics[width=0.8\textwidth]{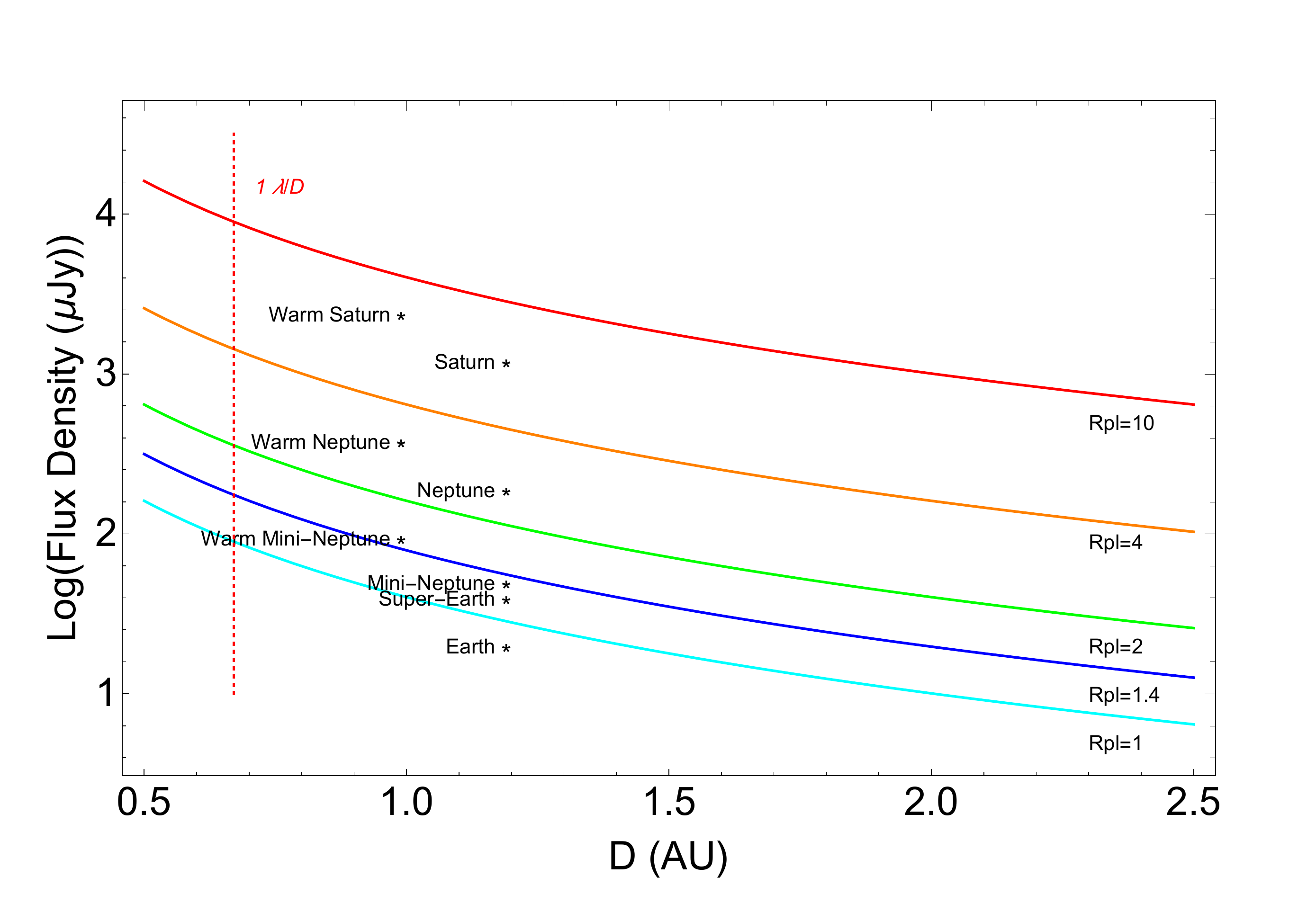}
\caption{The lines show the flux density at F1550C for planets of different radii (denoted in \rearth\ on the right) as a function of radial separation from \acen\ A based on a simple $T_{eff}\propto \textrm{D}^{0.5}$ relationship for an albedo of 0.3. Also shown are the predicted F1550C flux densities for the  detailed  models specified in Table~\ref{fluxes}. The dotted red vertical line shows the projected location of MIRI's $1\lambda/D=0.67$\arcsec\ Inner Working Angle at 15.5 $\mu$m. \label{brightness}}
\end{figure}

While it is unlikely that the MIRI observations discussed here will achieve the sensitivity needed to detect Earths or Super-Earths,  (1-2 \rearth), very long observations combined with new techniques of speckle suppression  may allow the detection of rocky planets. Thus, for completeness, we consider some scenarios for small planets (Figure~\ref{fig:flux}b), for example a 2 R$_\oplus$ ``water world''  for which   water is the dominant gas in the atmosphere.  A thick water cloud  forms  with the cloud  base  at 0.1 bar, and the top at $\sim$0.001 bar. Due to this thick cloud that extends to low pressures, the resulting spectrum is a black body at ~215 K. We also considered an Earth-like planet, with either 1  or 1.4 R$_\oplus$  (an Earth or  Super-Earth). We simulated the atmosphere using the standard, mid-latitude temperature-pressure profile  and the full photochemistry model developed in \citep{Hu2012}.  It is well known that thermal emission of Earth can be presented by a combination of cloud-free, low-altitude cloud, and high-altitude cloud atmospheres, e.g.  \citep{DesMarais2002,Turnbull2006}. But for simplicity, we assumed a cloud-free atmosphere noting that other cloud types have smaller thermal emission features. The emission spectrum is dominated by absorption of CO$_2$, H$_2$O, and O$_3$.

Finally, for subsequent analyses ($\S$\ref{dither},\ref{monte}), we  also used a simple blackbody relationship \citep{Traub2010}:

\begin{equation}
\label{bbody}
\teff= T_*\left(\frac{1-A}{4f}\right)^{1/4}  \left(\frac{R_*}{d}\right)^{1/2} = 275\,  L_*^{1/4}\, (1-A)^{1/4}\, {d}^{-1/2} \, K
\end{equation} 

\noindent where $L_*,T_*,R_*$ are the stellar luminosity, effective temperature and radius, $A$ the planet albedo, $d$ the planet's distance from the star in AU (Figure~\ref{brightness}), and $f=1$ is appropriate for full heat distribution. In the figure, the adopted albedo is 0.3, but in subsequent analyses, the albedo was drawn randomly between 0.15 and 0.65 appropriate to gaseous planets in our solar system, e.g. \citet{Cahoy2010}.

\section{Exozodiacal Dust Orbiting \acen\ A \label{HZdust}}

The zodiacal cloud and Kuiper belt in our  solar system have analogs in many other planetary systems.   The recently published HOSTS survey used the nulling interferometer of the Large Binocular Telescope (LBTI) to set preliminary  upper limits  of 26 times the solar system zodiacal  level for a sample of  solar type stars \citep{Ertel2018}. \citet{Wiegert2014} find suggestive, but hardly definitive evidence for a ring of cold dust (53 K) located at $\sim$70-105 AU around the \acen\ AB system at a level comparable to the Edgeworth-Kuiper belt in our own solar system \citep{Teplitz1999}. The HOSTS survey suggests that the level of warm zodiacal emission is  higher for stars associated with  cold dust emission detected at longer wavelengths by Spitzer or Herschel.

 \begin{figure}[h!]
 \centering
    \includegraphics[width=0.5\textwidth]{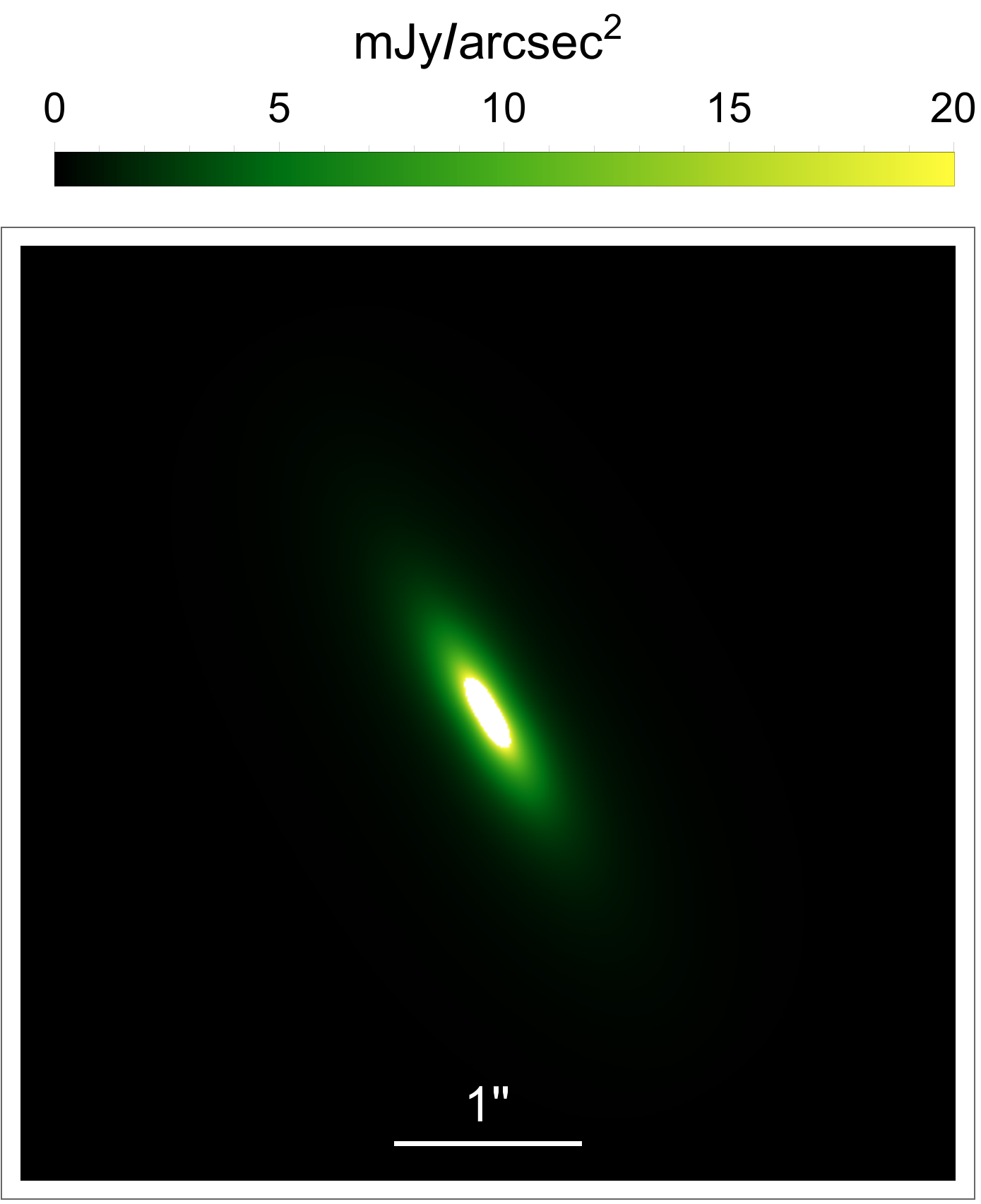}
   \caption{ A model of  a ``1 zodi" cloud seen around \acen\ A at 15.5 $\mu$m, generated using ZodiPic \citep{Kuchner2012} for a disk seen nearly edge-on (79$^o$). The image is 5\arcsec\ on a side. The total dust flux is 8.9 mJy, i.e about 10$^{-4}$ of the stellar flux at the same wavelength. \label{fig:Zodi}}
\end{figure}

The proximity of \acen\ means that JWST/MIRI can spatially resolve a warm zodiacal dust cloud without  an interferometer and thereby improve the detectability of the dust relative to purely photometric measurements \citep{Beichman2006a}. A model of a near-edge-on  ``1 zodi" cloud seen around \acen\ A at 15.5 $\mu$m can be generated using ZodiPic \citep{Kuchner2012}. Figure~\ref{fig:Zodi} shows an  5\arcsec$\times5$\arcsec\ image  of a cloud whose total dust flux density is 8.9 mJy, i.e about 10$^{-4}$ of the stellar flux at the same wavelength. Adopting an ``optimistic" HZ definition of 0.75 to 1.77 AU \citep{Kopparapu2017} for a sun-like star and correcting for stellar luminosity,  the excess flux density in the HZ is approximately 3 mJy, i.e. a total fractional excess of 3.6 x 10$^{-5}$ between 0.92 AU and 2.18 AU. This emission  would be spread over roughly 10 MIRI beams, or approximately 0.3 mJy per beam which is comparable in brightness to a ``Warm Neptune" (Table~\ref{fluxes}).  The detection of  emission at this level  ($\S$\ref{zodiobs}) is interesting for two reasons. First,  observing a spatially resolved excess in the Habitable Zone would be an important contribution to our knowledge of the evolution of exoplanet systems. Second, exozodiacal emission at the few Zodi level may set a limit to the size of  a HZ planet which might be detectable with MIRI's angular resolution \citep{Beichman2006b}.

Exactly how the exozodiacal dust is distributed is critical to its detectability and its effect on the detectability of any planets. Many exozodiacal clouds, e.g. Fomalhaut, HD69830, $\epsilon$ Eridani, have gaps rings or clumps often attributed to the presence of planets  \citep{Su2013, Beichman2005, Mawet2019} . A faint but homogeneous disk might simply be resolved away during the reference star subtraction while a clumpy cloud observed at the limit of JWST's angular resolution might be confused with one or more planets. Additional simulations and finally JWST observations will be required to assess these challenges.

\section{Overcoming the Observational Challenges}

The first challenge to finding one or more planets  orbiting \acen\ A is to select the preferred wavelength and instrument. Compared to NIRCam's   coronagraph operating at 4-5 $\mu$m with an Inner Working Angle (IWA) of 4-6 $\lambda/D$, MIRI's Four Quadrant Phase Mask (4QPM)  operating at  $\sim 1\, \lambda/D$  offers: comparable IWA,  improved  immunity to Wavefront Error (WFE) drifts and centering errors  \citep{Knight2012}, more favorable planet-star contrast ratio   at the expected planet temperatures at the IWA (200-300 K; Figure~\ref{fig:flux}), and lower brightness of  background stars. MIRI offers three 4QPM masks at 10.65, 11.4 and 15.5 $\mu$m. Although the shortest wavelength filter would have a smaller IWA, we have focused our discussion on  F1550C for a number of reasons: longer integration time before detector saturation\footnote{As calculated using the JWST exposure time tool. https://jwst.etc.stsci.edu/}  (10 sec vs 1 sec for F1550C vs F1065C), lower impact of wavefront drifts, good sensitivity across a broad range of planet temperatures (Figure~\ref{fig:flux}),  lower  confusion due to background stars, and complementarity to shorter-wavelength ground-based efforts ($\S$\ref{ground}).  

\subsection{Rejecting Starlight from $\alpha$ Cen A}

\begin{figure}[h!] 
\centering
\includegraphics[width=0.8\textwidth]{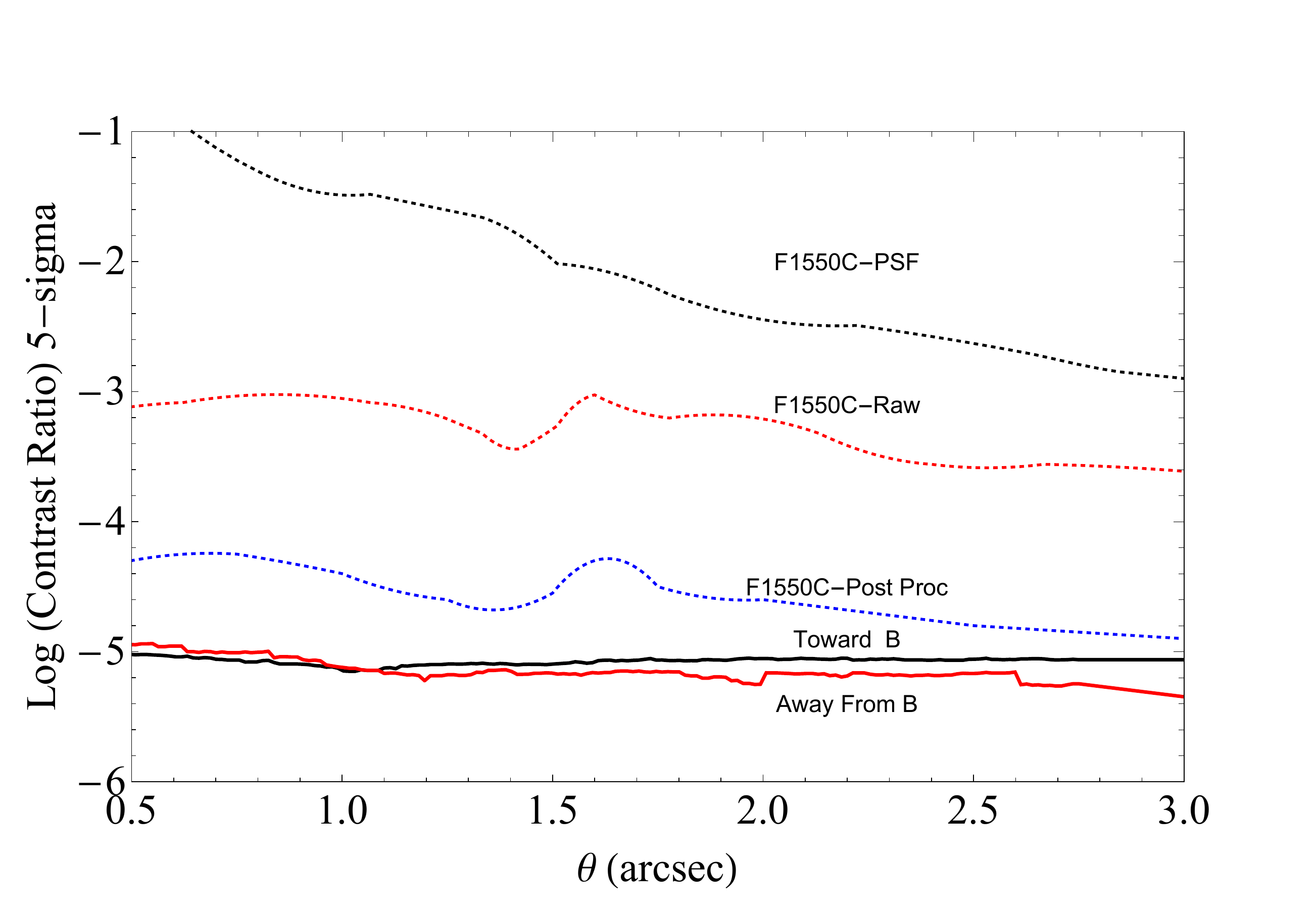}
\caption{Contrast curves for the F1550C curve:  PSF (dotted, black), Raw coronagraphic contrast (dotted, red), post PSF subtraction (dotted, blue)--all from \citep{Boccaletti2015}. The two solid curves show the contrast following our PCA post-processing with the upper black curve showing the influence in the direction of \acen\ B, located 7\arcsec\ away, and the lower red curve the contrast in directions away   from \acen\ B. The effect of \acen\ B is negligible with a few arcseconds of  \acen\ A. \label{fig:ContrastCurves}}
\end{figure}

MIRI's 4QPM reduces the central brightness  of a star by a factor of $\sim10^3$, operates as close to the star as 1 $\lambda/D\sim 0.48$\arcsec\ at $\lambda=15.5\, \mu$m (where $D$ is the telescope diameter),  and  achieves 10$^{-4}$-10$^{-5}$ rejection at a separation of 1\arcsec-2\arcsec\   using standard reference star subtraction (\citet{Boccaletti2015}, Figure~\ref{fig:ContrastCurves}).  As we discuss in $\S$\ref{dither} and show in the two lower lines in Figure~\ref{fig:ContrastCurves},  it should be  possible to improve on this performance with a specialized observing mode and advanced post-processing. 
 
 \subsection{Rejecting Starlight from $\alpha$ Cen B}

\begin{figure}[h!]
\includegraphics[height=0.8\textwidth]{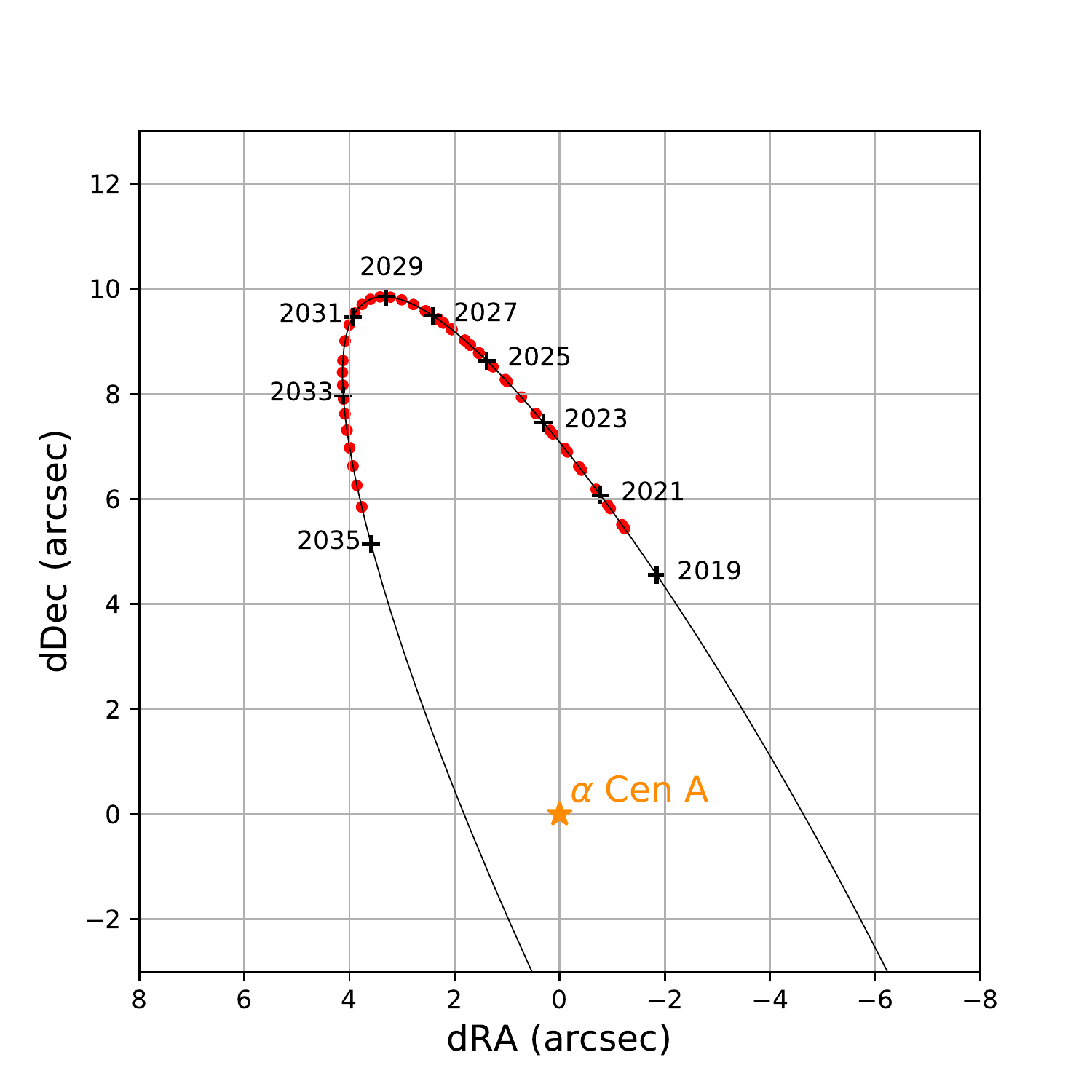}
\caption{ The orbit of \acen\ B around \acen A \citep{Kervella2016} is indicated with some possible observing dates ($>$2021) highlighted during the early years of JWST's operation. \label{orbit}}
\end{figure}

 Complicating the  issue is the presence of \acen\ B which will be located roughly 7-8\arcsec\ away from \acen\ B during the first few years  after JWST's launch. (Figure ~\ref{orbit}). We considered two methods for dealing with \acen\ B: 1) placing \acen\ B on  the transmission gap in the 4QPM \citep{Boccaletti2015, Danielski2018} to reduce the its central intensity at the cost of a limited selection of observing dates with the correct on-sky orientation; or 2)  optimize the target-reference star observations so as to minimize wave front error drifts while accepting   the deleterious effects of  the full brightness of \acen\ B falling on the detector.
 
A  positive aspect  of the 4QPM coronagraphic masks is the existence of  a gap, $\pm$3 pixels (at the half power points) or  $\sim$0.3\arcsec, located at the phase boundaries of the four quadrants. At these locations the transmission is reduced by a factor of $>$8 \citep{Danielski2018}.  There are semi-annual observing windows of a few days  duration during which \acen\ A can be centered behind  the coronagraphic mask while at the same time placing \acen\ B in one of the  gaps between adjacent  quadrants, thereby  reducing detector artifacts. However, as discussed in $\S$\ref{dither} this  approach  requires a non-optimized slew to a reference star which may induce changes in telescope's thermal environment resulting in  non-zero wavefront errors and a higher  level of residual speckles.

The alternative approach of placing \acen\ B in an unattenuated portion of the detector offers the advantage of a broader observing window at the cost of a greater risk of deleterious effects of the full intensity of \acen B. 

\begin{figure}[h!]
\centering
\includegraphics[width=0.8\textwidth,angle=0]{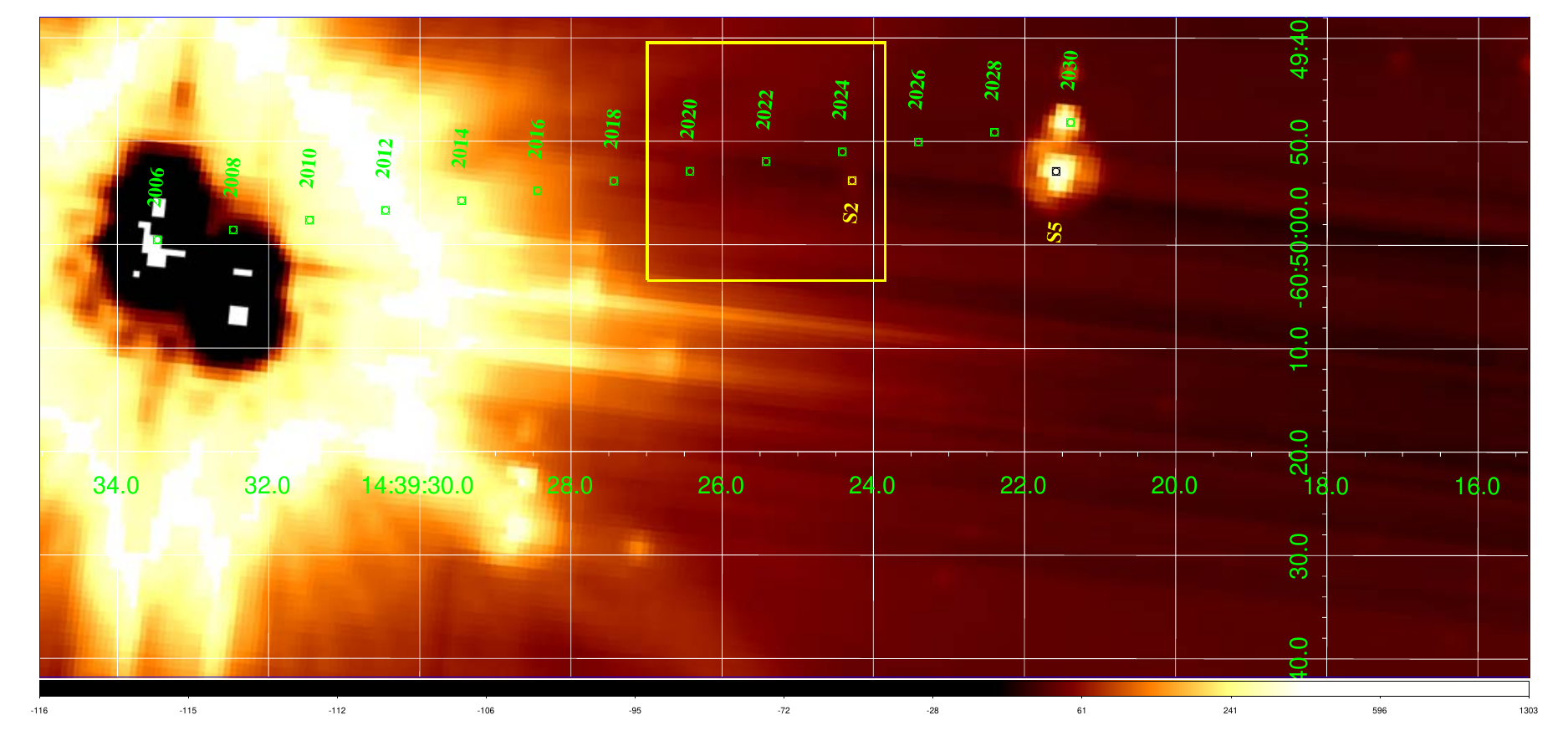}
\caption{ A Spitzer image (in celestial coordinates) of \acen\ AB \citep{Fazio2004}  taken in 2005 at  8.0  $\mu$m. The position of \acen\ A in 2022 is shown with a yellow square demarcating the approximate field of the 23\arcsec\ MIRI coronagraph. There are no Spitzer sources within the projected MIRI field at the level of a few mJy. The approximate position of \acen\ A is shown by  a series of green squares through 2030 when the source labelled ``S5" \citep{Kervella2016} approach \acen\ A itself. \label{Spitzer}} 
\end{figure}

\subsection{Confusion by Background  Stars and Galaxies}
 
The high proper motion of \acen\ ($\sim 3$\arcsec\ yr$^{-1}$ due West) means that  images from earlier epochs (Spitzer, HST, ground-based,  etc) can be  used to study the field where \acen\ will be during the JWST era and to identify background objects.  Figure~\ref{Spitzer} shows a 8 $\mu$m (Ch 4) Spitzer/IRAC image of \acen\ AB taken in 2005  \citep{Fazio2004} with the location  of MIRI's 23\arcsec\ coronagraphic field surrounding  \acen's projected position around  $\sim$2022. The brightest  stars in the vicinity are  $S2$ ($K_s=11.1$ mag)  which will pass within 1.6\arcsec\ of \acen\ A around 2023.4 and a brighter source $S5$  ($K_s=7.8$ mag) which will pass within 0.015\arcsec in 2028.4 \citep{Kervella2016}). \textit{The impending approach of $S2$ argues for observing \acen\ A  soon after launch to avoid the impact  of $S2$ on the observations}. 

\begin{figure}[h!]
\includegraphics[height=0.8\textwidth]{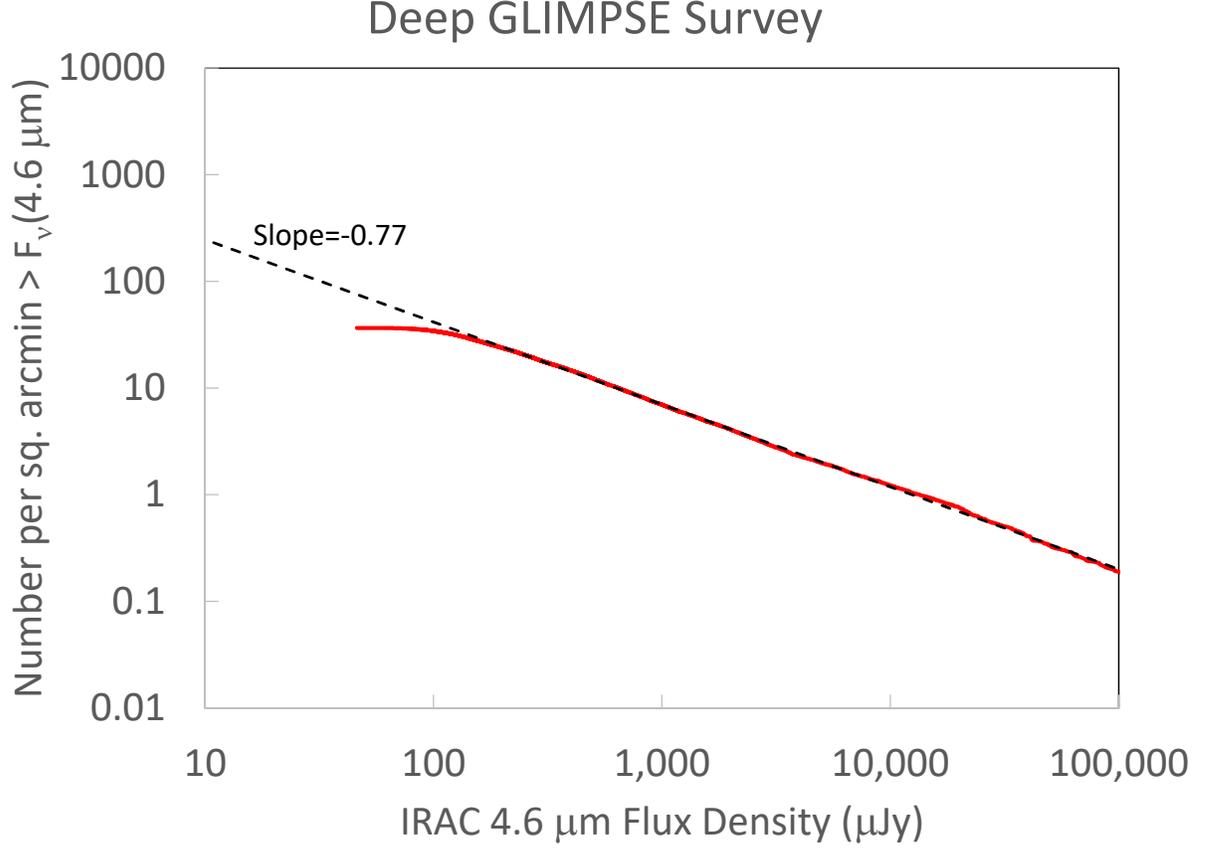}
\caption{Spitzer star counts at 4.6 $\mu$m  from the GLIMPSE survey at a position close to \acen\ are extrapolated below the confusion limit. The slope of the curve is typical of stellar populations in the Galactic Plane. We assume that background stars are fainter at the 15.5 $\mu$m wavelength of the MIRI coronagraph by a Rayleigh-Jeans factor of (15.5 $\mu$m/4.6 $\mu$m)$^2=11.3.$   \label{counts}}
\end{figure}%

Even if there are no obvious bright stars in the coronagraphic field it is important to estimate the level of contamination of background stars and galaxies at the expected levels of emission for our hoped-for planets, e.g.   F$_\nu$(F1550C)=20 $\mu$Jy for an Earth analog and  2.4 mJy for a warm Saturn (Table~1). To estimate the stellar background we take advantage of Spitzer's GLIMPSE survey  \citep{Churchwell2009} which covered a  region  near \acen.  We extracted from the Deep GLIMPSE catalog\footnote{https://irsa.ipac.caltech.edu/data/SPITZER/GLIMPSE/gator\_docs/GLIMPSE\_colDescriptions.html} sources  in a   $r=10^\prime$ region located at  galactic coordinates   (l,b)=(-315.3$^o$,-0.56$^o$),   just 0.5$^o$ away from \acen,  at   4.6 $\mu$m. The  4.6 $\mu$m data become confusion limited ($\sim$ 50 beams per source) at around  500  $\mu$Jy (Figure~\ref{counts}), but at fluxes brighter than this level   the plot of cumulative source counts, $N$, as a function of flux density, $S$,  $LogN(>S)/LogS$  has a  slope of $-0.77$, typical of a distribution of stars in the Galactic plane. 

If we extrapolate the source counts to lower fluxes  assuming that most of these objects have a Rayleigh-Jeans spectrum, then  we can estimate the number of background  sources expected within a $\pm$2.5\arcsec\ (3.3 AU) field  around \acen A at F1550C. The extrapolated number of 15.5 $\mu$m sources is 0.004 at  the  2.4 mJy brightness of a Saturn and 0.15 sources for a 20 $\mu$Jy  Earth (Table~1). Only at the brightness level of an Earth does the expected occurrence of background stars become a matter of concern,  while for a  Neptune the expected number of background sources in a $\pm$2.5\arcsec\ field is 0.03. The expected number of extra-galactic background sources is even lower at these flux levels. Using model sources counts from \citet{Cowley2018} we find that the  predicted number of galaxies at 15 $\mu$m within 2\arcsec\ of \acen A is less than 0.0035 at 20 $\mu$Jy. For host stars 10 to 20 times further away than \acen\ A,  the incidence of stellar and especially  extra-galactic background objects will be a much more serious problem.   Even though the stellar and extra-galactic  sources of false positives are rare, multi-color (F1065C vs F1550C)  and ultimately astrometric confirmation will be required to confidently reject background objects.

\subsection{Detector Performance Toward Bright Stars}

Stars as bright as \acen\ AB present unique challenges for the MIRI detector which is a 1024$\times$1024  arsenic-doped silicon (Si:As) IBC hybrid array \citep{Rieke2015, Ressler2015}. Even if placed behind one of the gaps in the 4QPM mask,  \acen\ B  would saturate portions of the detector and if not attenuated by a gap, the saturation problems would be even worse. To address  detector artifacts from very bright sources, we used an instrument testbed at JPL to conduct  tests on  an MIRI engineering model detector using an exact copy of the flight electronics. Appendix~\ref{AppendixMIRI} describes the test results in detail, but the primary conclusion is that  the tests reveal  no detector-based limitations to the detection of planets around the \acen\ A.

\section{Observational Scenarios}

The signal-to-noise (SNR)  of a detection near \acen\ A is driven by both photon noise due to unsuppressed starlight which can be  mitigated with increasing integration time and residual speckle noise which must be mitigated via improved PSF and speckle suppression.  
The envisioned technique of post-processing relies on the observation of a reference star with the small-grid dither technique. This technique compensates for possible jitter during the observation that slightly change the position of the target behind the coronagraph by artificially reproducing the same jitter effect while observing the reference star. 

Our simulations of the observational sequence show that we achieve a   reasonable balance between photon noise and residual speckle noise if we set the number integrations per dither point  to keep the ratio of {\it total} target to reference star observing time at 1:3. 
This ratio depends on the difference of magnitude between the target and the reference and the stability of the observations. In particular, it is a compromise between two extreme scenarios: 1) negligible level of jitter that would require a 1:9 ratio or 2) higher level of jitter that would allow a ratio closer to 1:1, assuming two stars of the same magnitude.
The adopted 1:3 ratio is a compromise that we would refine with further simulations and on-orbit information on the performance of JWST.

With this plan we can achieve  detections at the levels at the  $10^{-5}$ level at $>$1\arcsec\ as discussed below. An initial reconnaissance program sufficient to detect a 5$\sim$6 R$_\oplus$ planet would require  approximately 3.5 hours of on-target observing time.
 Adding in the $\sim3\times$ longer duration of reference star observation plus observatory overheads leads to a total $\sim$ 20 hr program according to the   JWST Exposure Time Calculator \footnote{https://jwst.etc.stsci.edu/}.

A single epoch of   F1550C  observations will produce a dataset which will both probe the limits of MIRI coronagraphy and result in either the detection of a planet or set limits at the $5\sim 6$ R$_\oplus$ level. MIRI might also detect solar system levels of exozodiacal emission ($\S$\ref{zodiobs}).  Subsequent observations at  multiple wavelengths would identify  background objects with stellar colors and provide   astrometric confirmation of detected objects.

\subsection{Reference Star Selection}
 
Coronagraphic imaging  to detect a 5 \rearth\ planet, not to mention  1  \rearth, presents  a daunting observational challenge. The choice of a reference star is critical  to removing the stellar point spread function (PSF) and residual speckles. To minimize observing time on the reference star and to maximize the level of speckle suppression it is important to  find  the best match in terms of brightness, spectral type and angular separation. Fortunately,  on the Rayleigh-Jeans tail of   photospheric emission,  color effects in  the narrow 6\% passbands of F1550C   filters are small compared with shorter wavelength observations.


There are  a number of options for reference star which also affect  the overall observing scenario. The closest reference to \acen\ A is, of course, \acen\ B. Using    \acen\ B has the advantages of minimal change in telescope configuration and  rapid target acquisition compared with  choosing a more distant reference star. The disadvantage is that one can never escape the influence of the $\sim$ 1 mag (at long wavelengths)  brighter \acen\ A to obtain a clean, uncontaminated PSF measurement.
Ground-based programs have adopted the \acen\ B approach using rapid chopping between the two stars ($\S$\ref{ground}). Here we examine a more conservative approach which takes a more widely separated,  single star to evaluate the PSF at the positions of both   \acen\ A and B. Interestingly, the two scenarios require roughly the same amount of wall clock time as determined by the JWST APT tool\footnote{http://www.stsci.edu/scientific-community/software/astronomers-proposal-tool-apt}, approximately 20 hours.

For  stars  as bright as [F1555C] $\sim$-1.4 mag, our choices are quite limited.  We used the IRAS Low Resolution Spectrometer Catalog \citep{LRS} to identify  potential reference stars: F$_\nu(12\mu$m) $>$ 50 Jy within 20$^o$ of \acen,  clean Rayleigh-Jeans photospheric emission, constant ratio ($<$10\%) of LRS brightness (F(\acen)/F(star)) across the F1550C band,  a low probability of variability during the 300 day IRAS mission ($VAR<15$\%), and no bright companions within 100\arcsec. Table~\ref{refstars} lists  potential reference stars. The ratio of the LRS spectra of the (unresolved)  \acen\ AB system to these stars is constant across the F1550C  bandpass to $<$ 1\%. 

\begin{center}
\begin{tabular}{lccc}
\multicolumn{4}{c}{Table 4. Candidate Reference Stars\label{refstars}} \\
Star & Spec Type&Sep (deg)&[(12$\mu$m)] mag$^1$\\
\hline
BL Cru&M4/5 III&17&-0.93 \\
 BO Mus&M6II/III&15&-1.7 \\
DL Cha&M6III&18&-0.64 \\
V996 Cen &Carbon Star&8&-0.70 \\
$\epsilon$ Mus &M4III&17&-2.09 \\
del01 Aps&M4III LPV&20&-1.36 \\
$\zeta$ Ara&K3III&19&-1.17 \\
\hline
\acen\ B & K1V &0.002&-0.6$^2$\\ \hline
\multicolumn{4}{l}{$^1$Magnitude from IRAS Catalog; $^2$ estimated from shorter wavelengths} \\
\end{tabular}
\end{center}


\subsection{Achieving Highest Imaging Contrast  \label{dither}}

\begin{figure}[t!] 
\centering
\includegraphics[width=0.5\textwidth]{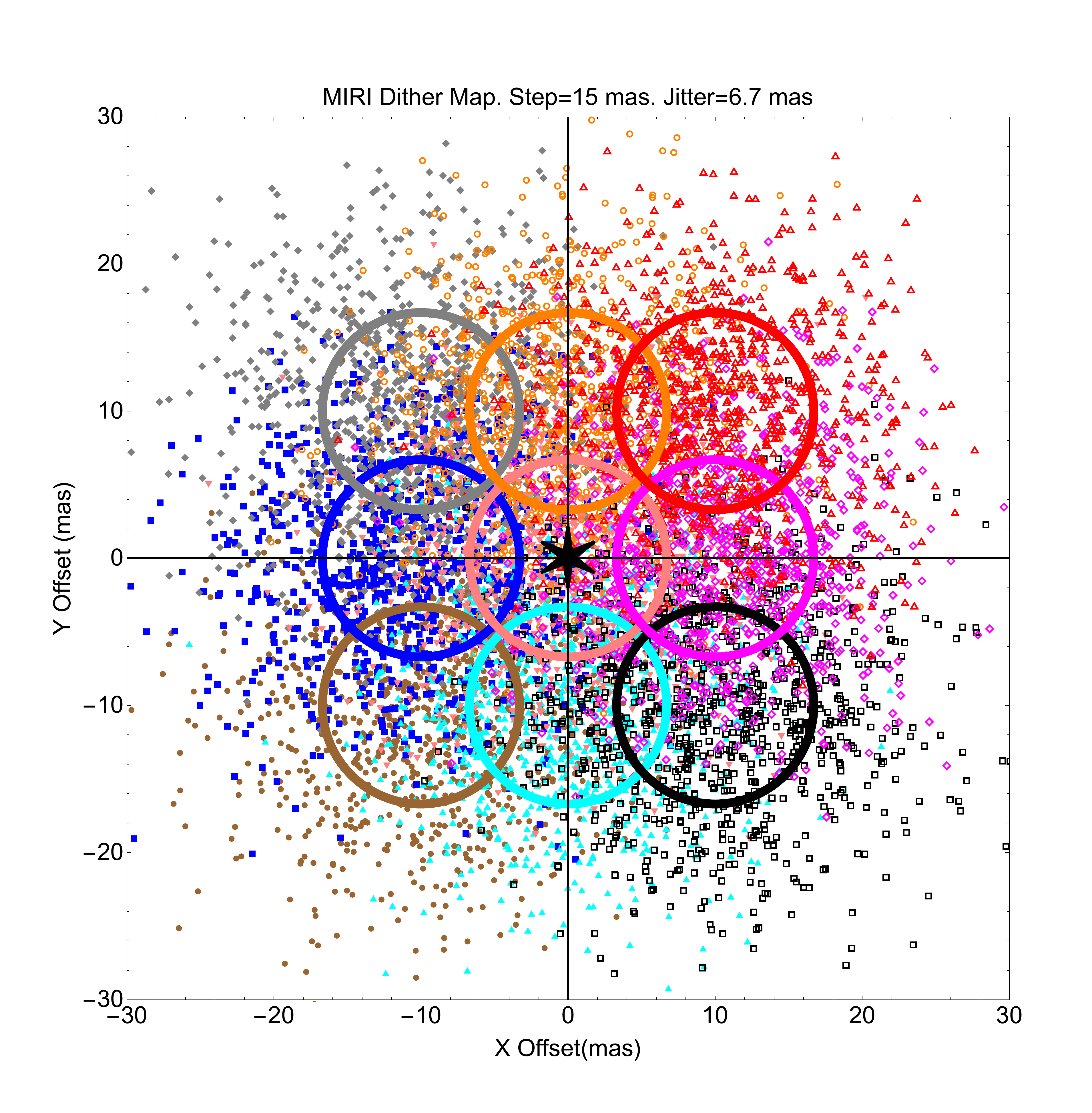}
\caption{The 9 point grid dither observation strategy combined with the diversity added by the 6.7 mas jitter  of the telescope (denoted by the circles) during acquisition, allows for enhanced diversity in the reference images (each denoted by a small symbol) to be used for reduction.\label{fig:RefAcq}}
\end{figure}

Achieving the  sensitivity needed  to detect  planets requires  aggressive post-processing techniques  to reduce the residual  speckles from both \acen\ A and B. We have simulated an observing scenario which places a reference star at the positions of both \acen\ A and B. The small grid 9-point dither pattern available for MIRI observations is used at the position of \acen\ A. The 15 mas micro-steps in the dither pattern   combined with  the 6.7  mas pointing jitter during the observation \footnote{https://jwst-docs.stsci.edu/display/JTI/JWST+Pointing+Performance}  improve the sampling of the point spread function (PSF) and thus the ability  to remove stellar speckles (Figure~\ref{fig:RefAcq}).  We used a Principal Component Analysis (PCA) Algorithm~\citep{Soummer2012,Amara2012} to generate a sequence of reference images using  all the individual short-exposure frames    obtained during the observations. 


For each  image we generated a wavefront map realization which differed from its predecessor by a random amount and by a linear drift as described by \citet{Perrin2018} and which will be described in more detail below. The resultant wavefront maps were   used to create two PSFs using  the IDL version of {\it 
MIRImSIM}\footnote{https://jwst.fr/wp/?p=30}: the on-axis PSF representing \acen A  and an off-axis PSF at 7\arcsec\ representing \acen B at its projected separation in $\sim$2022.   For this simulation we generated  468 exposures (52 separate pointings each with a 9 point dither pattern) for reference star at the position of A and 100 pointings (with no dither)  for the reference star at the position of B. These individual reference star images were combined to generate a PSF library with  25,000 individual  images (out of a possible 46,800) of the \acen\ AB system.

We also generated 200 images of \acen\ A  including  planets  of different sizes and  locations (1 to 10 \rearth, 0.5-3\arcsec). We also generated over 450 reference star images ($\S$\ref{WFE}). On orbit we will obtain many more images by using short exposures, $\sim$10 sec, to avoid saturation at the core of \acen\ A  and to further increase image diversity\footnote{The ETC shows that the wings of the unattenuated \acen\ B are not saturated beyond 1\arcsec-2\arcsec\ in 10 sec.}.  Experimenting with the PCA reductions showed that windowing the images around \acen\ A to a 5\arcsec$\times$5\arcsec~enhanced the performance of the PSF subtraction. Indeed, given that the region of interest does not include the region where the center of \acen\ B falls, excluding this region avoids the bias that \acen\ B induces in the reference PSF computation with PCA. 

Although nominal values for  readout noise, photon noise from the sky and  telescope background  \citep{Ressler2008,Rieke2015, Boccaletti2015}   were added to the images, the signal from the planet itself and/or  speckle noise from \acen\  A dominate the measurement within $\sim$3\arcsec. The final  image had a total integration time of 3.5 hr and was obtained by combining the short exposure frames for \acen\  A, \acen\ B and one of the simulated planets.

\begin{figure} 
\centering
\includegraphics[width=0.9\textwidth]{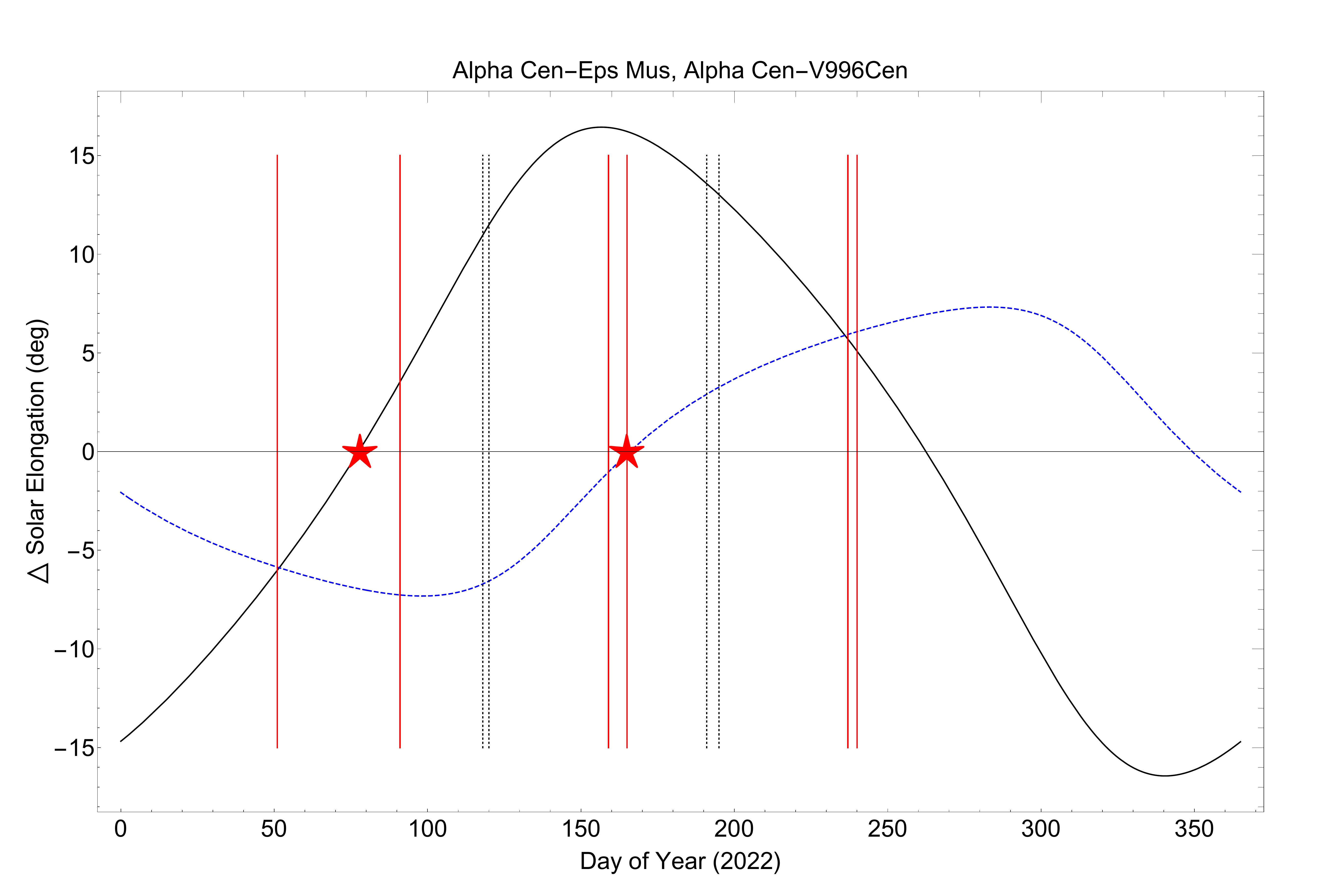}
\caption{The curves show the difference between the solar elongation angles between \acen\ A and two possible reference stars (Table~\ref{refstars}), $\epsilon$ Mus (solid black line) and V996 Cen  (dotted blue line), through the course of one year, nominally 2022. Pairs of vertical red bars show times  when \acen\ B can be located  within a 4QPM quadrant while the pairs of dotted black bars show times when \acen\ B can be hidden behind one of the 4QPM gaps. Minimizing the change in solar elongation angle during a slew between  \acen A and either star is possible on select  days marked by red stars. The periods where \acen\ B can be placed behind the gap result in  slews with large changes  in solar elongation angle, 5$^o$-10$^o$, between the target and reference stars.
\label{solar}}
\end{figure}

 \subsubsection{Minimizing the   Effects of Wavefront Drift \label{WFE}}
 
The ability to detect faint companions is dominated by the stability of the nominal 132 nm of wavefront error (WFE) of the JWST telescope. According to \citet{Perrin2018}, a slow-varying thermal WFE ranging from 2 to 10 nm can be expected depending on the change in solar elongation (and thus in the telescope's thermal balance). Assuming a minimal solar elongation difference as illustrated in Figure~\ref{solar}, we adopted a slow-varying thermal WFE of 2 nm RMS over the total observation of either αCen A or the reference star. The wavefront changes were distributed across small-, medium-, and large spatial scales following the prescription of \citet{Lightsey2018, Perrin2018}. For a scenario requiring a large change of solar elongation angle $\geq10^o$, we used  initial WFE  maps  for the reference and target stars which differed from one another by a random  2-10 nm. 

We simulated two different scenarios of wavefront evolution (Figure~\ref{solar}). In one case \acen\ B was located behind one of the 4QPM gaps while in the other  \acen\ B was located at 45 degrees relative to the 4QPM boundaries. Those two scenarios have  different implications for the observations. Putting \acen\ B   on one of the gaps  attenuates the star \citep{Boccaletti2015, Danielski2018} with a positive effect on the level of speckles and photon noise on the final image. However, this option requires a stricter time constraint that limits our ability to optimize the solar elongation difference between the target and its reference star. Thermal models of telescope performance show that large changes in elongation angle produce sudden WFE drifts. These sudden WFEs lead to a higher level of residual speckles, which proves to be very detrimental to sensitivity. Positioning  \acen\ B  in between two quadrant boundaries relaxes this time constraint and enables us to optimize the difference in solar elongation.  In the first scenario with \acen\ B  on one of the gaps, the difference in solar elongation is estimated to $\sim10^o$, which could result in a wavefront offset  between 2-10 nm RMS between target and reference star WFE distributions \citep{Perrin2018} whereas with  \acen\ B falling  between two quadrant boundaries, the difference in solar elongation can be reduced to near zero  which \citep{Perrin2018} suggests would result in a slowly evolving wavefront difference of 2 nm RMS or less.

\begin{figure}[h!] 
\centering
\plottwo{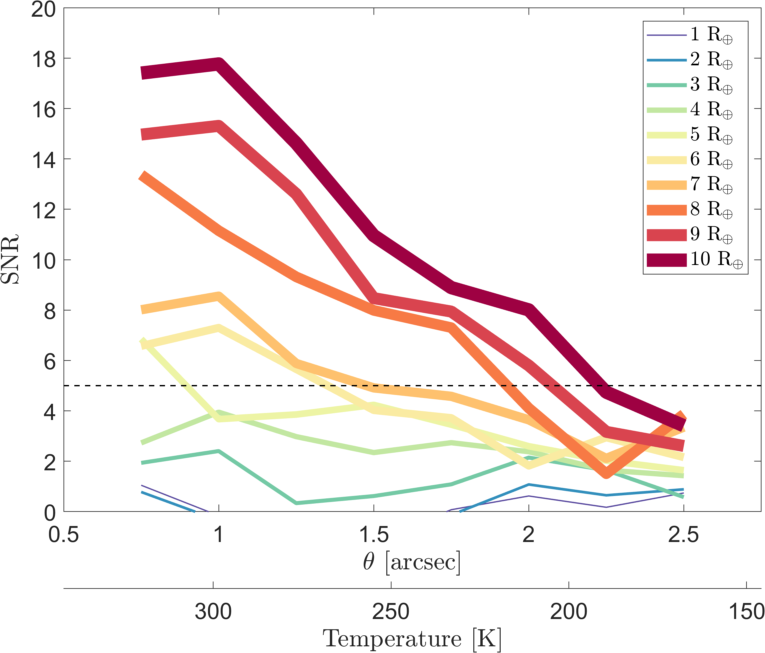}{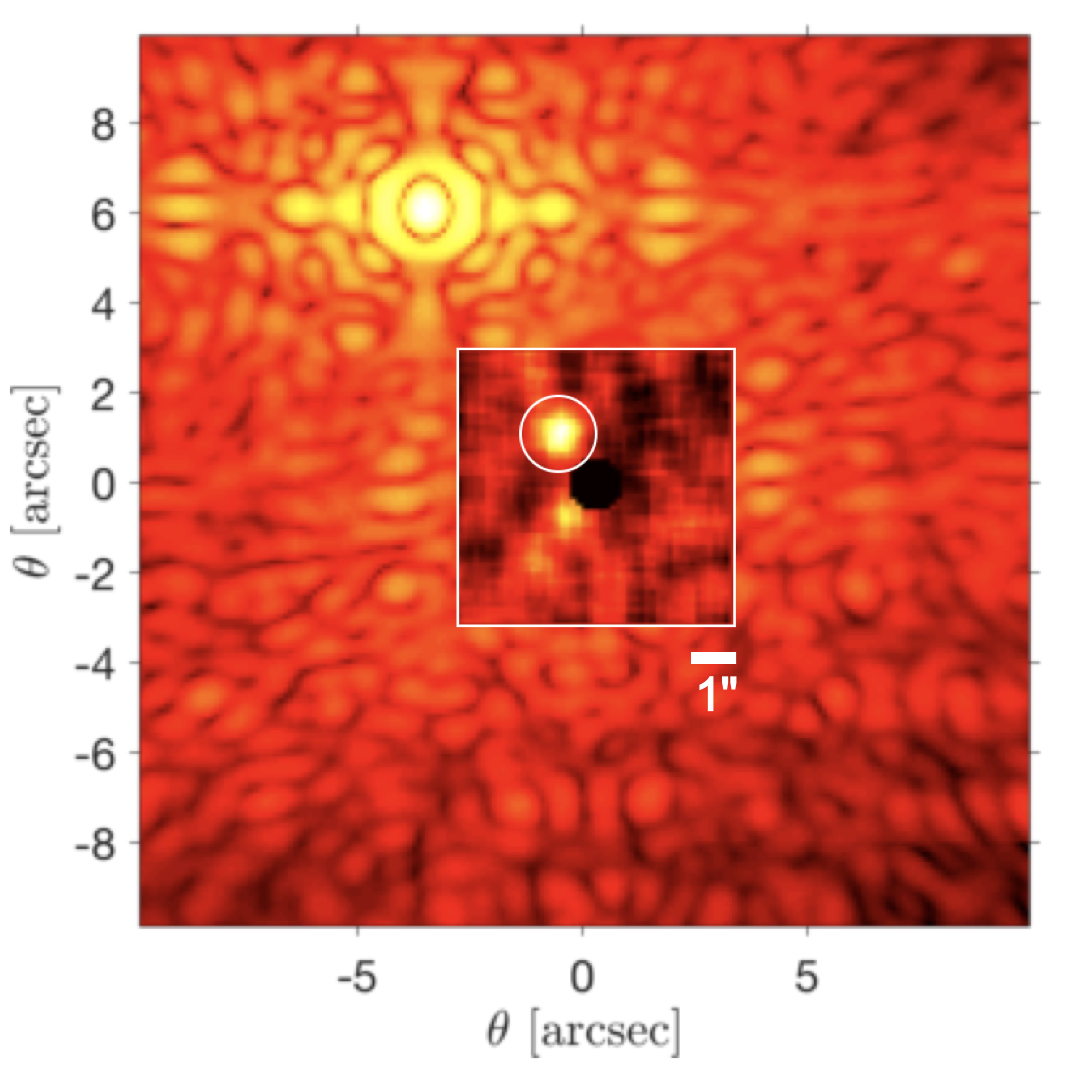}
\caption{a,left) The sensitivities for different planet sizes at the expected angular separation range of detection were computed for a slow thermal varying wavefront error of 2 nm RMS (\textit{left}). b,right) Simulation  for the 2 nm case (left) for the \acen\  system with the F1550C filter centered on \acen\ A, with \acen\ B on the top left, 7 arcseconds away. The PCA reduction of the data is done on a 5\arcsec$\times$5\arcsec\  central portion of the full image (white square, the scales inside the square are different from outside). A 10 \rearth\  planet is detected at 1.5\arcsec\ (white circle). \label{fig:SNR}}
\end{figure}

Figure~\ref{fig:SNR} compares the signal to noise ratio (SNR) in the PCA-processed images for different planet radii and temperatures (separations) for  wavefront errors of 2 nm RMS (left). 
The noise at each radial offset  was determined by taking the median of the values within an 1 $\lambda/D$ annulus at that radius. The SNR drops for smaller planet radius and with increasing star-planet separation due to the decrease in planet temperature. The effective limit (SNR$\sim$5) of these observations is roughly 5-6 \rearth\  within 1.5\arcsec.  The 10 nm case (not shown) is even less favorable, strongly favoring observing scenarios which minimize WFE drifts.  Figure~\ref{fig:SNR}b shows a final F1550C image showing both \acen\ B and an inset showing the PCA-corrected region with a 10 \rearth\  planet at 1.5\arcsec\ from \acen\ A.


Our simulations show that the scenario where  \acen\ B falls on one of the gaps, the change in wavefront stability resulting from large changes solar angle greatly offsets the advantage of lower \acen\ B intensity.  The scenario where \acen\ B falls within a   quadrant is more favorable to the detection of small planets.    

These results reinforce the fact that, in the present case of direct imaging of exoplanets around \acen, but also for more general cases for direct imaging of circumstellar environments, optimizing the wavefront stability through the adequate choice of reference star and optimization of observing times is crucial. On a separate note, observing sources off the gap  allows observations with the (Angular Differential,  ADI) strategy via rolls during a given visit or via multiple visits.

Figure~\ref{fig:ContrastCurves} shows that in the present era, when the separation between the two stars is $\sim$ 7\arcsec, the presence of \acen\ B has a relatively small  effect on the ability to detect a planet orbiting \acen\ A. Not until 1.5\arcsec\ does \acen\ B appear to have a significant effect on post-processed contrast ratio, increasing from 5$\times10^{-5}$ to  8$\times10^{-5}$  on the \acen\ B facing side.

Finally, we assessed the effect of  increasing the integration time {\it within a single visit} by a factor of 2 or more and did not see any  improvement in the detectability of smaller planets. Our analysis suggests that the noise floor is set by residual speckle noise, not photon noise. Furthermore, within a given visit, the range of roll angles is modest, $\pm5^o$, so that  the power of ADI is limited. The maximum   10$^o$ roll results in only a two pixel shift at 1.5\arcsec, compared with the   0.6\arcsec\ resolution at 15.5 $\mu$m. However, combining multiple visits with a broader range of angles and independent samples of the WFE map and drift, should produce improved sensitivity to small planets. Such visits will be necessary in any event to ensure that any planets obscured within the IWA are observed. Repeating this basic 3.5 hr observing block described here 9 times with independent wavefront realizations could result in a  three-fold improvement in sensitivity and  allow detections of planets down to $\sim$3 \rearth.

\section{Detecting and Imaging the exozodiacal Cloud \label{zodiobs}}

\begin{figure}[h!]
 \centering
    \includegraphics[width=1.\textwidth]{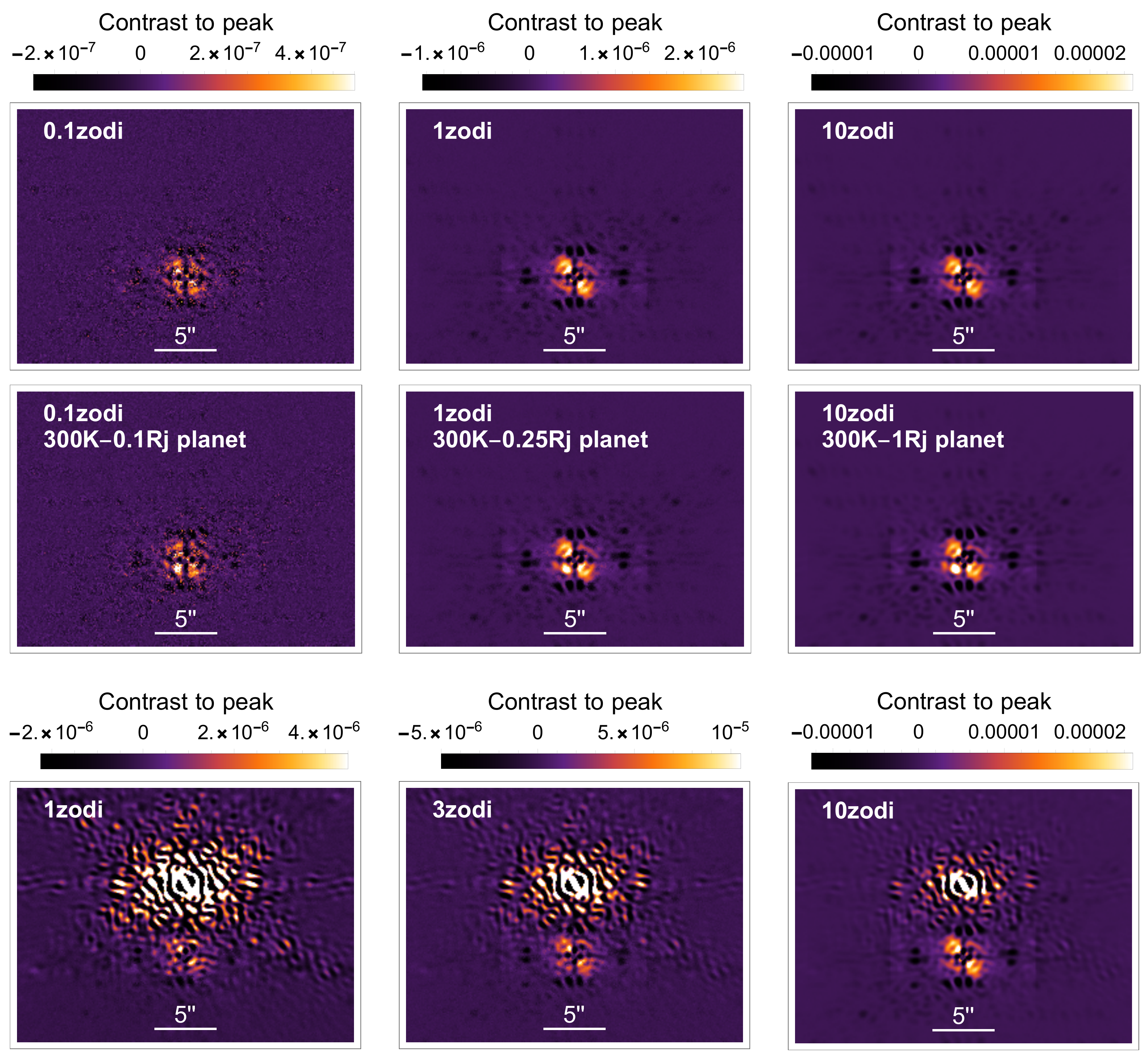}
\caption{ The zodi model (Figure~\ref{fig:Zodi}) as observed with MIRI using the 4QPM mask in a 10 hr exposure. The top two rows show PCA reductions of observations ignoring   the influence of  \acen\ B. The  data were  taken using Small Grid Dithers for models with  three different levels of zodiacal emission (0.1, 1, 10) Zodi, without and with a 300 K planet of three different radii (0.1, 0.25, and 1) R$_{Jup}$. The bottom panel adds in the effect of \acen\ B for the no planet case. \label{fig:ZodiObs}}
\end{figure}

Observations of the ZodiPic model ($\S$\ref{HZdust}, Figure~\ref{fig:Zodi}) have been simulated using the observing scenario described above  and were reduced  using PCA analysis with the results shown in Figure~\ref{fig:ZodiObs}. The figure shows the result of a 10 hour exposure. The resolved exozodiacal cloud is readily detectable at levels above $\sim$3 Zodi (or $\sim$5 in a single 3.5 hr exposure) and the  excess integrated around the entire Habitable  Zone  would probably be detectable below that level. Detection of a Habitable Zone exozodiacal dust cloud at this level would be a unique contribution by JWST to our knowledge of the environment of the Habitable Zone of a solar type star.

\section{Probability of Detecting a Planet Around \acen\ A \label{monte}}
  
We use a Monte Carlo analysis \citep{Beichman2010} to assess the probability of finding a planet  of a given radius, R$_p$ (\rearth), and semi-major axis, SMA  (AU), in the F1550C filter.  The flux density of the planet is calculated from the blackbody function (Eqn~\ref{bbody}) at the appropriate planet radius and orbital location, $d$.  Figure~\ref{brightness} shows the range of planet brightness which approaches a few mJy for 10\rearth\ planets. For simplicity we have assumed complete redistribution of absorbed stellar energy so that there is no day-night temperature gradient and no difference in temperature as a function of phase angle.   Figure~\ref{fig:flux} shows that  a simple blackbody  (Figure~\ref{brightness}) over-estimates  the brightness of Earth analogs with  a deep CO$_2$ absorption feature at 15 $\mu$m. Such planets are already far below the JWST detection limit considered here, so the absorption figure was ignored in the Monte Carlo calculation. 
 
An  input population   is randomly drawn from the sample described by Eqn~\ref{occurrence}   ($\S$\ref{prospects})  with the additional constraint of  a Radial Velocity cut of  5 m s$^{-1}$ appropriate to a 100\mearth\ planet at 2 AU \citep{Zhao2017}. Orbital eccentricity is  randomly drawn between $0<$ eccentricity$<$ 0.5. To convert from planet mass to the planet radius needed to estimate thermal emission, we follow \citet{Wolfgang2016}  and  adopt  $M=C(R/R_\oplus)^\gamma$ with values for $C$ and $\gamma$  from their Table 1: C=1.6 \mearth\ and $\gamma=1.8$. Similarly, we take the  dispersion around the predicted radius is taken from their Eqn. (3), $\sigma=\sqrt{\sigma_1^2+\beta (R/R_\oplus -1)}$ with $\sigma_1=2.9$\mearth\  and $\beta$=1.5.  

In the simulation planets are  placed  at randomized locations in their orbits. Planets with apoastron greater than 3 AU are excluded due to stability arguments.  An apoapse of 3 AU is  used as a hard limit, because there appear to be no islands of stability beyond  that   distance (Figure~\ref{stability}, \citet{Quarles2016}). The planets are confined to the plane of the \acen\ AB binary system \citep{Kervella2016} with  an added dispersion in the inclination of 5$^o$. Each planet is started on its orbit at a random time of periastron passage so that the Monte Carlo analysis  samples all possible positions of planets relative to the  IWA  of the MIRI coronagraph. This analysis adopts the transmission of the 4QPM mask \citep{Boccaletti2015} and the one dimensional coronagraph performance curve shown in Figure~\ref{fig:ContrastCurves} which is based on the PCA post-processing ($\S$\ref{dither}).

Figure \ref{yield}a shows contours of the probability of detecting a planet of a given radius and semi-major axis in a single  visit with 3.5 hours of on-target integration time. There is a broad plateau of detectability $\sim$50\% for R$>$5R$_\oplus$ and $1<$ SMA$<2$ AU.  Figure~\ref{yield}b shows  detectability contours based purely on photometric considerations, i.e. ignoring  geometrical constraint due to planets being obscured within the IWA, and show what planets might be detected in the limit of multiple visits.

Figure~\ref{yield} does not take into account the restriction on planets due to the RV observations. Figure~\ref{smooth}a shows a smoothed histogram of all detected planets, similar to Figure~\ref{yield}b, while Figure~\ref{smooth}b shows the distribution of  planets which could be detected and still be consistent with the $\sim$5 m s$^{-1}$ PRV upper limit. Using the \citet{Fernandes2019} occurrence rates,  Eqn~\ref{occurrence}, the fraction of all planets detectable  within  the 5 m s$^{-1}$ RV limit and a 5 \rearth\ MIRI limit is only 5\%. A more extensive campaign of multiple visits (with independent wavefront realizations) could push to lower radii and higher completeness ($\S$\ref{incomplete}). A 3 \rearth\ MIRI limit could detect $\sim13$\% of all of the  planets expected on the basis of the (poorly) known planet population and consistent with the RV limit; however, as noted in $\S$\ref{prospects}, the occurrence rates \citep{Fernandes2019}   could  be a factor of 3 lower in a binary system \citep{Kraus2016}. 

\subsection{Sources of Incompleteness \label{incomplete}}

\begin{figure}
\centering
\plottwo{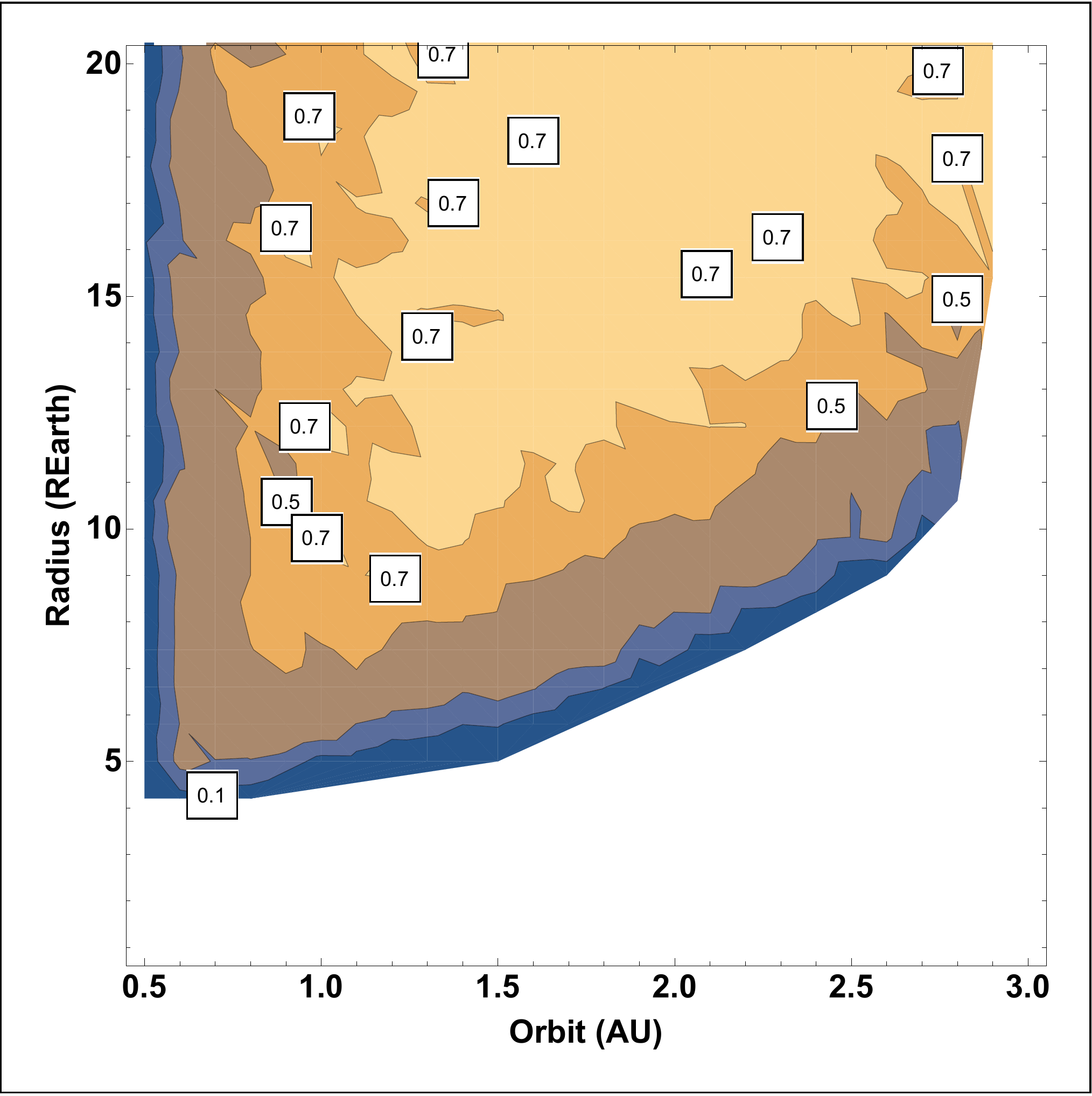}{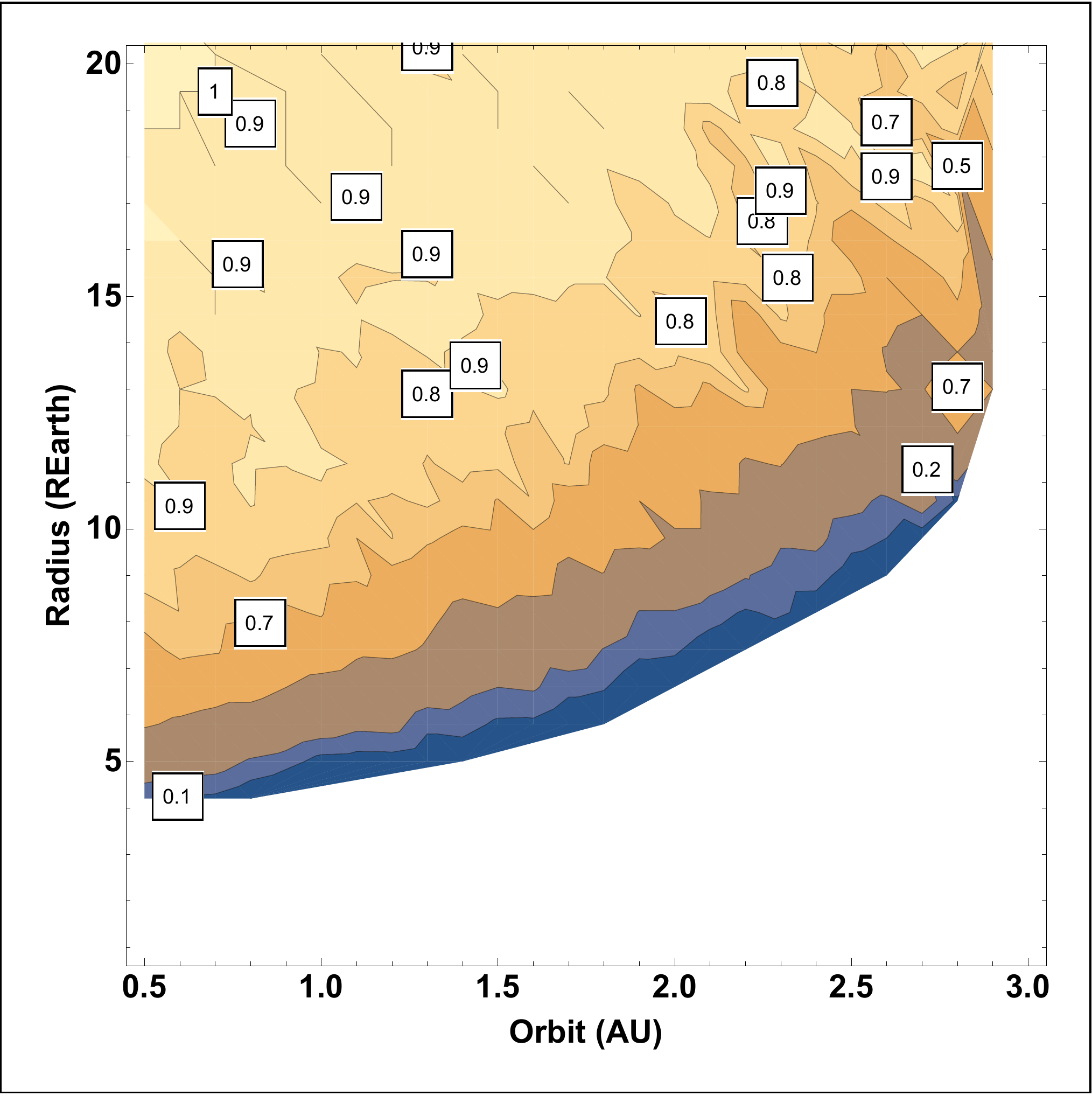}
\caption{a, left) A plot showing the detectablity of planets with a specified radius and semi-major axis (SMA) in a single visit, averaged over  ranges of albedo, orbital eccentricity and orientation as described in the text. The contour levels show the fraction of planets detected in a given (Radius, SMA) bin.  b, right) same plot but showing sensitivity-limited detectability which ignores geometrical incompleteness due to a planet being hidden within the Inner working Angle. \label{yield}}
\end{figure}

\begin{figure}
\centering
\includegraphics[width=0.7\textwidth]{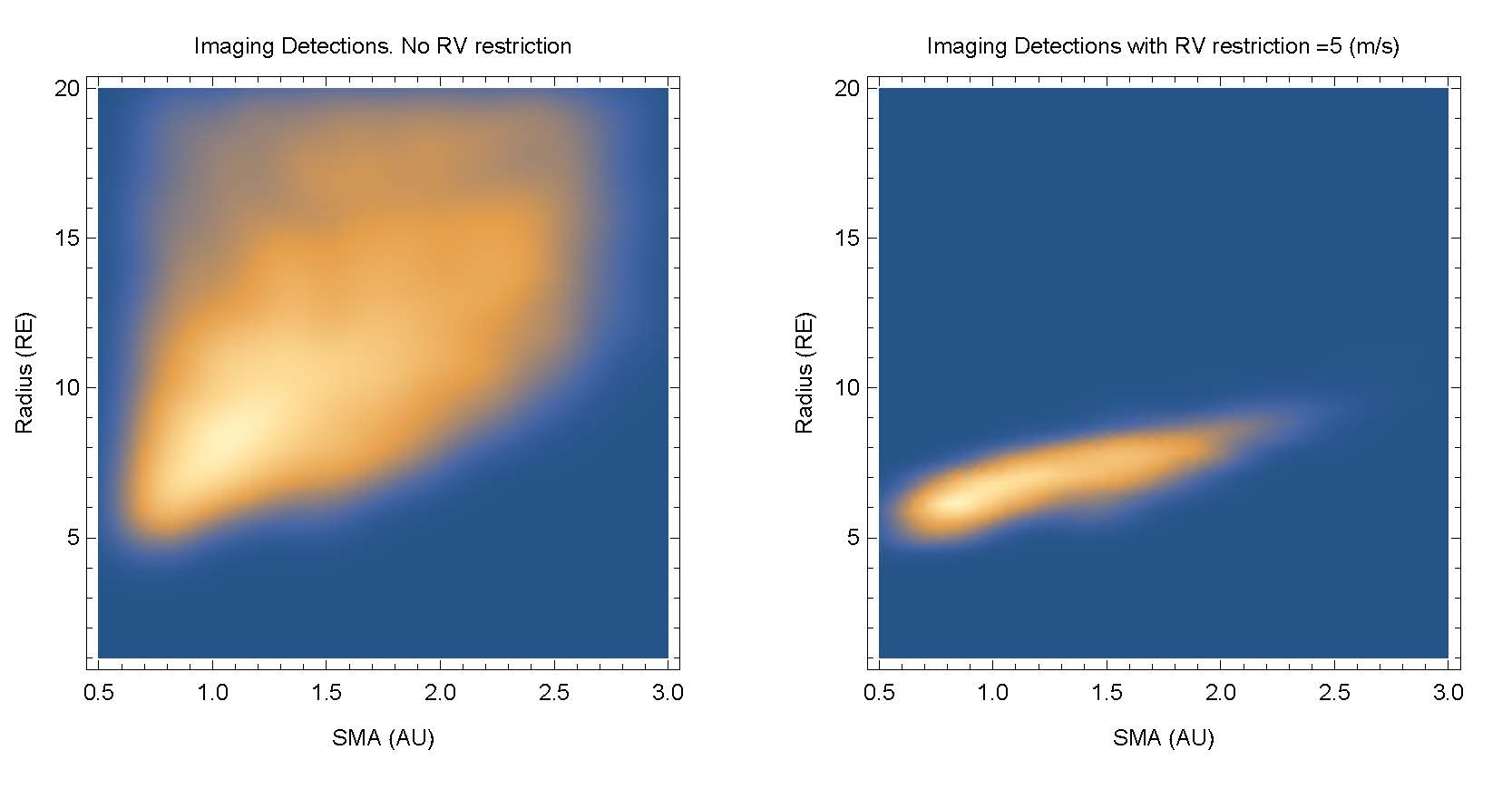}
\caption{a,left) The locus of all potentially detectable planets in (Radius-SMA) space similar to Figure~\ref{yield}b. b, right) the locus of all detected planets subject to the RV limit of 5 m s$^{-1}$. The intensity scale is arbitrary. \label{smooth}}
\end{figure}

Because \acen\ A is seen close to edge on, a planet can be  missed because its semi-major axis (or apoastron for an eccentric orbit) never takes it outside the Inner Working Angle of the coronagraph or simply  not far enough to be in a region of reduced speckle noise. Thus, the IWA and the contrast limit close to the IWA limit  the semi-major axis at which planets can be detected. Second, planets with orbits larger than the IWA can still be missed as they pass  behind the IWA in their orbit.  Thus, the maximum fractional  detectability  for a planet at SMA=1.2 AU (0.9\arcsec) with respect to the IWA of 0.49\arcsec\ at F1550C is $1-\frac{2}{\pi} ArcSin(\frac{IWA}{SMA})$=63\% in a single visit. As planets move further out, the fraction of time they are missed for geometrical reasons decreases. But, as they move further out, their temperature drops so they might be missed for reasons of low  SNR. These two effects account for the general shape of the detectability in Figure~\ref{yield}a.   The solution to the problems of  geometrical incompleteness is carrying out multiple observations over a number of epochs as pointed out in many studies of this question \citep{Brown2005, Brown2015}.

Two additional sources of incompleteness are not accounted for in Figure~\ref{yield}. First is the increased noise level in the direction of \acen\ B and second from the possibility that at any one instant a planet may hide behind  one of the 4QPM's quadrant gaps. Figure~\ref{fig:ContrastCurves} shows remarkably little difference in the post-processing curves in the direction of \acen\ B relative to other directions within the region of interest, $<3$ AU. \acen\ A is simply overpowering at these separations relative to \acen\ B located 7-8\arcsec\ away.

\begin{figure}[h!]
\centering
\includegraphics[width=0.7\textwidth]{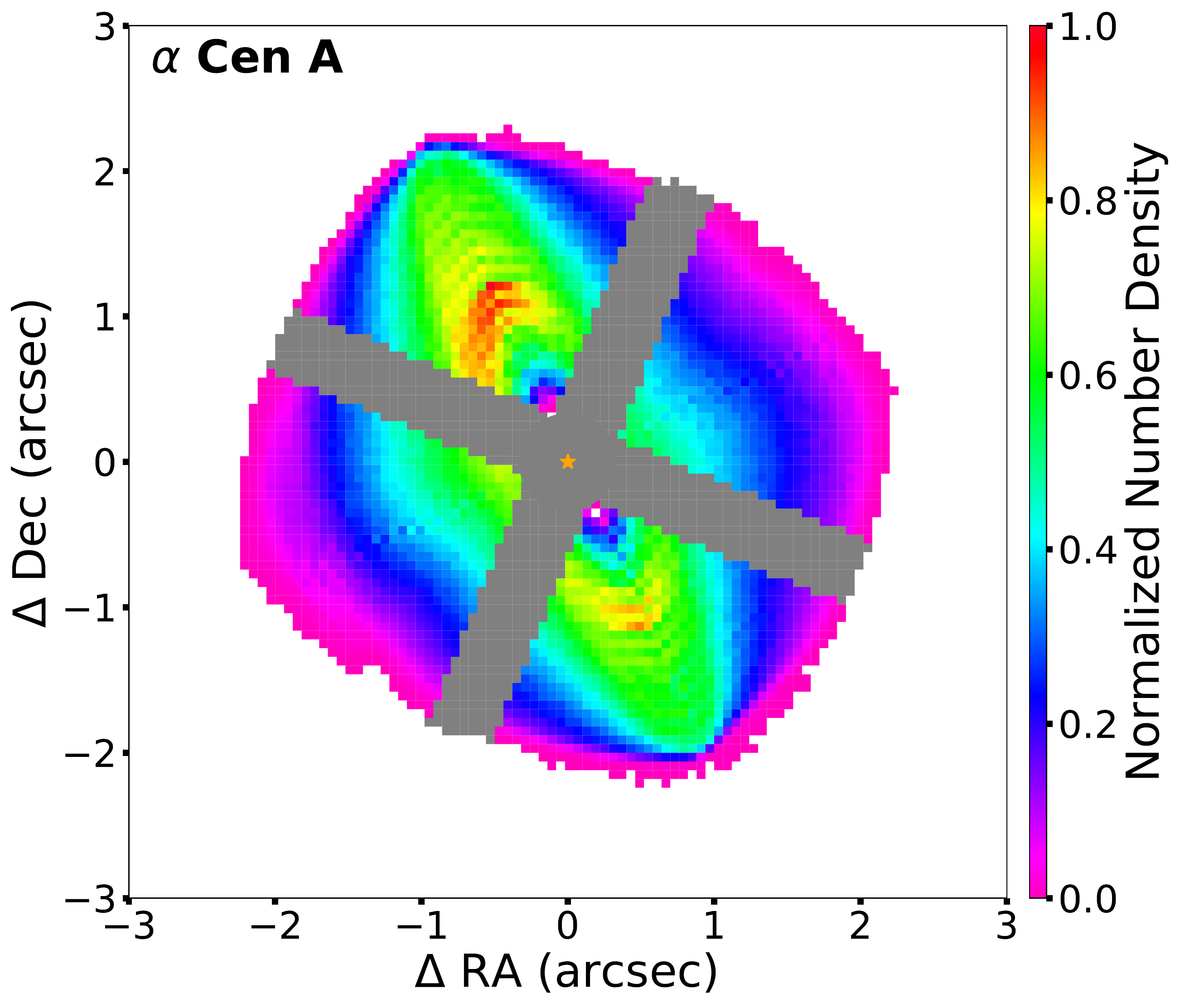}
\caption{Projection of initial conditions that are stable on $10^5$ yr timescales onto the sky plane. The stable initial conditions are binned for resolution at 15.5 $\mu$m, where the color scale denotes a normalized number density of stable initial conditions within each bin.  Bins that do not contain any stable initial conditions are colored white.  The regions defined by the 4QPM mask and the gaps between adjacent quadrants are plotted in gray over the region of potential planet stability. These regions block roughly 24\% of the coronagraphic field outside of the IWA. \label{4QPMgap}}
\end{figure}

The second source of incompleteness not taken into account in Figure~\ref{yield} are  the dead areas defined by the 4QPM gaps. We test the second source by performing numerical simulations using a modified version of the \texttt{mercury6} integration package designed to evolve planetary orbits in binary systems \citep{Chambers2002}.  These simulations use the orbital solution from \cite{Pourbaix2016} for the binary orbit and evaluate the stability on $10^5$ yr timescale for Earth-mass planets over a range of initial SMAs (1--3 AU), eccentricity vectors ($e_p\cos\omega_p \leq 0.9$, $e_p\cos\omega_p \leq 0.9$), and mutual inclinations ($<90^\circ$).  Figure~\ref{4QPMgap} shows the projection of initial conditions that are stable (survive for 10$^5$ yr) and binned using the expected angular resolution ($\sim$0.3 \arcsec) at 15.5 $\mu$m to identify a normalized number density of potential orbits on the sky plane (see color scale).  Projecting the gap width onto the $\sim$2.5 AU zone of stability \citep{Quarles2016} reveals that the incompleteness due to the gaps outside of the IWA is around 24\%. This source of incompleteness can be mitigated by multiple visits at different orientations. Aligning the gaps with the \acen\ AB axis results in an incompleteness of 60\%--another reason to avoid this  observing scenario.

\subsection{Comparison with Ground-based Initiatives\label{ground}}

The proximity  of the \acen\ system makes it a compelling target for ground-based studies in the N (10 $\mu$m) band despite the high sky background. As described in \citet{Kasper2017} and \citet{Kaufl2018}, the European Southern Observatory (ESO)  in collaboration with the Breakthrough Initiative has   modified the VLT mid-IR imager VISIR to  enhance its ability to search for potentially habitable planets around both components of \acen. The  NEAR (New Earths in the Alpha Cen Region)  concept combines adaptive optics using the deformable secondary mirror at UT4, a new vector vortex coronagraph \citep{Mawet2005} optimized for the most sensitive spectral bandpass in the N-band, and fast chopping for noise filtering.

The recently demonstrated sensitivity of the NEAR instrument is 650 $\mu$Jy (5$\sigma$ in 1 hour, \citet{Kaufl2018}). Assuming no systematic errors intervene,  a 100 hr observing program with NEAR could have the sensitivity to detect a 2 \rearth\  planet with an Earth-like emission spectrum  at $\sim 3 \lambda/$D$\sim1$AU and a temperature around 300 K. This result, if achieved, could complement JWST's MIRI search by extending inward to smaller, hotter planets. In the long term, the NEAR  experiment is relevant for the Extremely Large Telescope/METIS instrument \citep{Quanz2015} which would benefit from the telescope diameter ($D$),  $D^1$ gain in inner working angle and the $D^4$ gain in photometric sensitivity  due to the ELT's  39 meter aperture.

 Dynamical searches for planets orbiting \acen A are continuing. The new generation of PRV instruments such as ESPRESSO \citep{Gonzalez2018} should be able measure down to a few Earth masses, although the presence of \acen\ B presents observational challenges  at  binary separations smaller than a few arcseconds. On the other hand,  both the ALMA and the VLT Gravity interferometers are taking advantage of this binarity by searching  for a planet-induced astrometric wobble in the separation between \acen\ A and B at millimeter \citep{Akeson2019} and near-IR wavelengths \citep{Gravity}, respectively. Dynamical detections from any of these techniques would add critical information on the mass and orbit of any planet found via  direct imaged--whether from JWST or other experiments now underway.

\section{Conclusions}

With careful observation planning and advanced post-processing techniques JWST's MIRI coronagraph could detect planets  as small as $5$ \rearth\ at 15.5 $\mu$m in a single $\sim$20 hr visit (combining $\sim$ 3.5 hr of on-target integration plus reference star and other overheads). Multiple visits would enhance completeness, provide astrometric confirmation, and push to still lower planet radii.  These additional observations would also help to refine orbital data and open a search for additional planets. Detection at MIRI wavelengths would  lead to an estimate of the planet's effective temperature and thus its radius which would depend only weakly on the assumed albedo. Of course, the actual performance of JWST in terms of wavefront error and especially WFE stability remains unknown as does the performance of its detectors. A more sustained campaign could push this radius limit down to  $\sim$3 \rearth.

MIRI could also detect an exozodiacal dust cloud at the level of $3\sim5\times$ the brightness of our own cloud. Depending on the strength and distribution of the exozodiacal dust, such emission could mask the light of any planet.

JWST data, in conjunction with   ground-based observations would provide refined characterization of any detected planets: PRV measurements  with both current and next generation instruments such as CHIRON and ESPRESSO \citep{Zhao2017, Gonzalez2018} would yield a refined orbit and the planet's mass from which we would  determine its bulk composition; VLT/NEAR detections at shorter wavelengths, $\sim$ 10 $\mu$m, would refine the spectral energy distribution. Ultimately, instruments combining high contrast imaging with high spectral resolution spectroscopy  on 30-40 m telescopes would open up the prospect of exoplanet spectroscopy of a planet orbiting in  the Habitable Zone  of a solar type star \citep{Snellen2015,Wang2017}.

\section{Acknowledgements}

Some of the research described in this publication was carried out in part at the Jet Propulsion Laboratory, California Institute of Technology, under a contract with the National Aeronautics and Space Administration.
Copyright 2019 California Inst of Technology. All rights reserved.

\begin{appendix}  
 \section{MIRI Detector tests\label{AppendixMIRI}}
 
\begin{figure}
    \centering
    \includegraphics[width=0.8\textwidth]{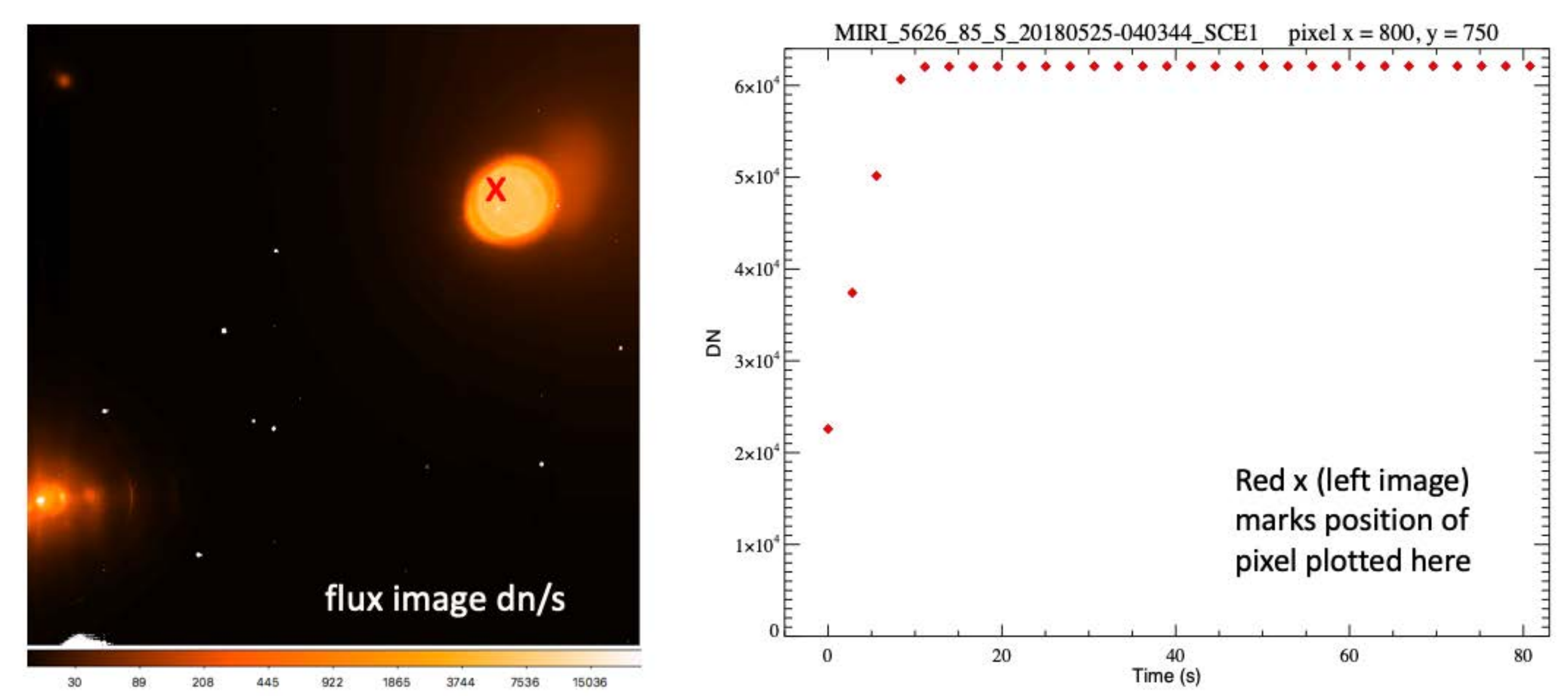}
\caption{Left) Full frame  MIRI engineering model detector. The
tests includes two sources from a masked black body, a faint point like object in the top left, and an extended disc structure in the top right. An unfocussed LED source can be seen in the bottom left of the image. The units of the image are flux (Data Numbers, DN/s) as calculated from the slope of unsaturated frames. Right) the signal recorded from one pixel in the disc black body source.}
\label{detector1}
\end{figure}

Estimates based on the JWST Exposure Time Calculator (ETC\footnote{https://jwst.etc.stsci.edu/}) for the  unattenuated signal from \acen\ B is  approximately $5\times10^7$ electrons s$^{-1}$ at 15.5 $\mu$m which is well above the saturation limit for the MIRI detectors. To explore the implications of such a bright source in the focal plane we carried out a series of tests using the flight-like configuration at JPL. Figure~\ref{detector1} shows the resulting image from the bright target test. The test  setup  was set so that it would quickly saturate the detector with a signal of a factor of 10 more than the saturation limit. Figure~\ref{detector1}  plots the signal recorded for one
illuminated pixel for each of the 30 groups in one integration. The test source saturated the detector in 4 groups or approximately 11 seconds for a total time per integration of 80 seconds. The total exposure time of 10 integrations was 14 minutes.

\begin{figure}
    \centering
    \includegraphics[width=0.8\textwidth]{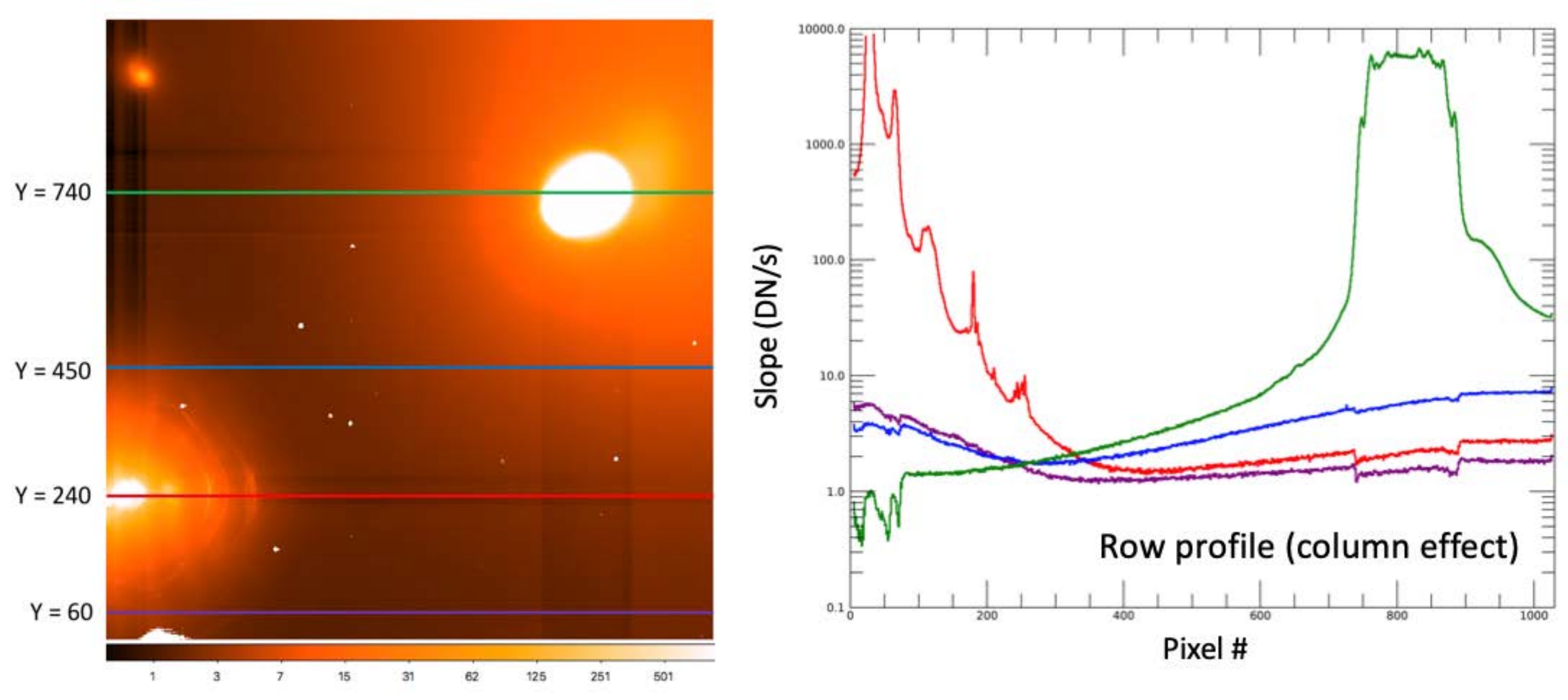}
\caption{a, left)  the same test data as Figure~\ref{detector1}   but with the scale set to highlight faint structure in the background. b. Colored lines crossing the image horizontally mark rows whose intensities are shown on the right. b, right) The profiles of the marked rows in the detector showing the effects of the column effect in the rows underneath of brightest source.}
\label{detector2}
\end{figure}

The resulting test image (Figure~\ref{detector1}) shows a good detection of the 3 sources used in the test despite the ``super" saturation of the detector. Other than glints and optical effects which originate in the test bench setup, there is no significant impact in the image quality from the ``super-saturation" of the detector. In Figure~\ref{detector2} we set the scale of the test image to enhance the background to reveal  faint structures associated with the rows and columns of the detected sources. Row profiles of the image, also presented in Figure~\ref{detector2}, show that the background artifacts are 4 orders of magnitude lower in flux than the sources in the image. 

\begin{figure}
    \centering
    \includegraphics[width=0.8\textwidth]{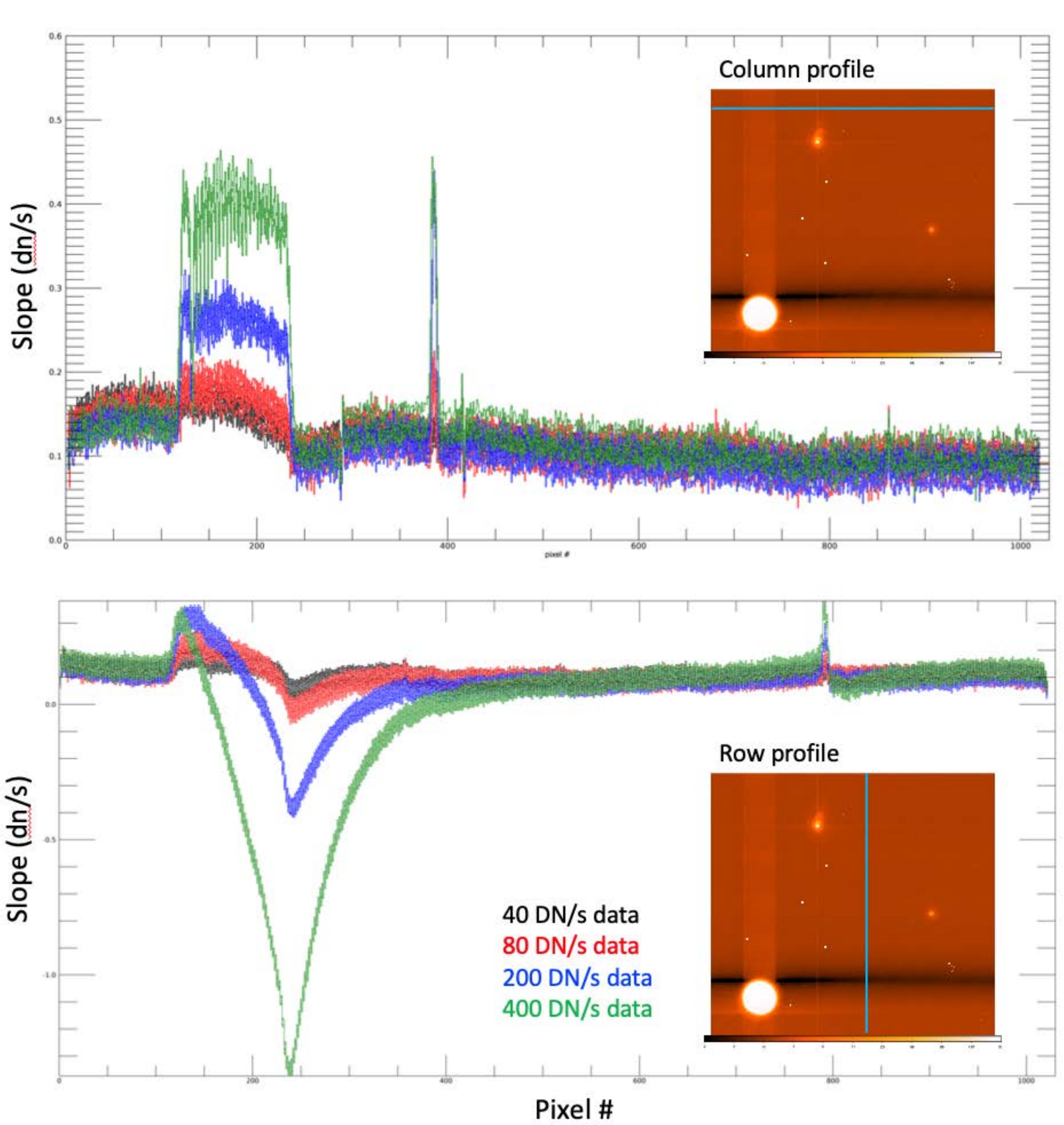}
\caption{JPL MIRI test data results showing data with 3 black body sources at four different flux levels. Row (bottom) and
column(top) profiles at each flux level highlight the extent of the artifacts in each direction.}
\label{detector3}
\end{figure}

This row and column structure in the detectors had been previously identified by the MIRI test team and is associated with bright source detections (Figure~\ref{detector3}). We believe the artifacts will have little impact on the detection of planets the following reasons. First, the row and column artifacts are accentuated in the JPL test images due to the very low backgrounds in the test conditions - a factor of three less than we expect at the shortest MIRI wavelength range of 5.7 microns.  For the higher backgrounds expected from the MIRI 4QPMs (F1550C) the column and row effects will be significantly diluted. (Figure~\ref{detector3}) also shows that, although the effect is flux dependent, it is limited to the columns in which there are bright
sources, therefore the effect from \acen\  B should be limited to the columns in which it is placed. 

However, the row effect extends beyond the source rows, in the read direction up the detector, with a dependence on source size. Therefore, there is a possibility that the row artifact could affect planet detection if \acen\ B  were placed in a lower quadrant of the 4QPM. However, we expect the point like nature of  \acen\ B  will help reduce the amplitude of this artifact. Lastly, the artifacts have shown to be highly uniform in amplitude in the row and column direction, therefore preliminary efforts to correct the image based on median column and row filtering have proved promising. In summary, we find no limitations from the point of view of MIRI detectors to the detection of planets around the \acen\ AB system from MIRI ground detector testing. Including the case of “super” saturation, which is expected in the observation of the \acen\ AB system with the MIRI 4QPMs.

\end{appendix}

\end{document}